\documentclass[twocolumn]{aastex63}

\usepackage{natbib}
\usepackage{graphicx}
\usepackage{epstopdf}
\usepackage{amsmath}
\usepackage{amssymb}
\usepackage{bm}
\usepackage{color}
\usepackage[normalem]{ulem}

\def\gs{\mathrel{\raise0.35ex\hbox{$\scriptstyle >$}\kern-0.6em
\lower0.40ex\hbox{{$\scriptstyle \sim$}}}}
\def\ls{\mathrel{\raise0.35ex\hbox{$\scriptstyle <$}\kern-0.6em
\lower0.40ex\hbox{{$\scriptstyle \sim$}}}}

\newcommand{\hei}{He\,{\sc i}}
\newcommand{\feii}{Fe\,{\sc ii}}

\newcommand{\civ}{C\,{\sc iv}}
\newcommand{\oiii}{[O\,{\sc iii}]}

\newcommand{\Hb}{H$\beta$}
\newcommand{\La}{Ly$\alpha$}

\newcommand{\dhb}{\mathcal{D}_{{\rm H}\beta}}
\newcommand{\dciv}{\mathcal{D}_{{\rm C IV}}}
\newcommand{\dlya}{\mathcal{D}_{{\rm Ly}\alpha}}
\newcommand{\hbm}{{\rm H}\beta}
\newcommand{\civm}{{\rm CIV}}
\newcommand{\lam}{{\rm Ly}\alpha}
\newcommand{\ignore}[1]{}

\shorttitle{AGN STORM~XII.\ Broad-Line Region Modeling of NGC\,5548}
\shortauthors{Williams et al.}

\begin{document}

\title{\Large Space Telescope and Optical Reverberation Mapping Project.\\
XII.\ Broad-Line Region Modeling of NGC 5548}

\correspondingauthor{Peter R. Williams}
\email{pwilliams@astro.ucla.edu}


\author[0000-0002-4645-6578]{P.~R.~Williams}
\affiliation{\ignore{UCLA}Department of Physics and Astronomy, University of California, Los Angeles, CA 90095, USA}

\author[0000-0003-1065-5046]{A.~Pancoast}
\affiliation{\ignore{CfA}Harvard-Smithsonian Center for Astrophysics, 60 Garden Street, Cambridge, MA 02138, USA}

\author[0000-0002-8460-0390]{T.~Treu}
\altaffiliation{Packard Fellow}
\affiliation{\ignore{UCLA}Department of Physics and Astronomy, University of California, Los Angeles, CA 90095, USA}

\author{B.~J.~Brewer}
\affiliation{Department of Statistics, The University of Auckland, Private Bag 92019, Auckland 1142,
New Zealand}

\author[0000-0001-6481-5397]{B.~M.~Peterson}
\affiliation{Department of Astronomy, The Ohio State University, 140 W 18th Ave, Columbus, OH 43210, USA}
\affiliation{Center for Cosmology and AstroParticle Physics, The Ohio State University, 191 West Woodruff Ave, Columbus, OH 43210, USA}
\affiliation{Space Telescope Science Institute, 3700 San Martin Drive, Baltimore, MD 21218, USA}

\author[0000-0002-3026-0562]{A.~J.~Barth}
\affiliation{\ignore{UCI}Department of Physics and Astronomy, 4129 Frederick Reines Hall, University of California, Irvine, CA 92697, USA}

\author[0000-0001-6919-1237]{M.~A.~Malkan}
\affiliation{\ignore{UCLA}Department of Physics and Astronomy, University of California, Los Angeles, CA 90095, USA}

\author[0000-0003-3242-7052]{G.~De~Rosa}
\affiliation{Space Telescope Science Institute, 3700 San Martin Drive, Baltimore, MD 21218, USA}

\author[0000-0003-1728-0304]{Keith~Horne}
\affiliation{\ignore{SUPA}SUPA Physics and Astronomy, University of St. Andrews, Fife, KY16 9SS Scotland, UK}

\author[0000-0002-2180-8266]{G.~A.~Kriss}
\affiliation{Space Telescope Science Institute, 3700 San Martin Drive, Baltimore, MD 21218, USA}


% ...

\begin{abstract}
We present geometric and dynamical modeling of the broad line region for the multi-wavelength reverberation mapping campaign focused on NGC 5548 in 2014.
The dataset includes photometric and spectroscopic monitoring in the optical and ultraviolet, covering the \Hb, \civ, and \La\ broad emission lines.
We find an extended disk-like \Hb\ BLR with a mixture of near-circular and outflowing gas trajectories, while the \civ\ and \La\ BLRs are much less extended and resemble shell-like structures.
There is clear radial structure in the BLR, with \civ\ and \La\ emission arising at smaller radii than the \Hb\ emission.
Using the three lines, we make three independent black hole mass measurements, all of which are consistent.
Combining these results gives a joint inference of $\log_{10}(M_{\rm BH}/M_\odot) = 7.64^{+0.21}_{-0.18}$.
We examine the effect of using the $V$ band instead of the UV continuum light curve on the results and find a size difference that is consistent with the measured UV-optical time lag, but the other structural and kinematic parameters remain unchanged, suggesting that the $V$ band is a suitable proxy for the ionizing continuum when exploring the BLR structure and kinematics.
Finally, we compare the \Hb\ results to similar models of data obtained in 2008 when the AGN was at a lower luminosity state.
We find that the size of the emitting region increased during this time period, but the geometry and black hole mass remain unchanged, which confirms that the BLR kinematics suitably gauge the gravitational field of the central black hole.
\end{abstract}



\section{Introduction}

Broad emission lines in active galactic nuclei (AGN) are thought to arise from the photoionization of gas in a region surrounding a central supermassive black hole.
The geometry and dynamics of this so-called broad line region (BLR), however, are not well understood.
Since a typical BLR is only on the order of light days in radius, this region nearly always cannot be resolved even in the most nearby AGN, with rare exceptions (e.g., 3C 273, \citealt{Sturm++18}).
Emission-line profiles can provide some information about the line-of-sight (LOS) motions of the gas, but more data are required to extract the BLR structure and dynamics.

The technique of reverberation mapping \citep{blandford82, peterson93, peterson14, ferrarese05} utilizes the time lag between continuum fluctuations and emission line fluctuations to extract a characteristic size of the BLR.
Paired with a velocity measured from the emission-line profile, these data provide black hole mass measurements to within a factor, $f$.
This factor, of order unity, accounts for the unknown BLR structure and dynamics.
Velocity-resolved reverberation mapping takes this one step further by breaking up the line profile into velocity bins and studying how each part responds to the continuum.
This method has found results that are consistent with gas in elliptical orbits for some objects, while others indicate either inflowing or outflowing gas trajectories \citep[e.g.,][]{bentz09, denney09c, barth11, barth11b, du16, agnstorm5}.
With a similar goal, the code {\sc MEMEcho} \citep{horne91, horne94} has been used to recover the response function, which describes how continuum fluctuations map to emission line fluctuations in LOS velocity$-$time-delay space.
Comparing these velocity-delay maps to those produced by various BLR models has pointed towards a similar range of BLR geometries and dynamics \citep[e.g.,][]{bentz10b, grier13a}.

In this work, we utilize an approach to directly model reverberation mapping data using simplified models of the BLR, first discussed by \citet{pancoast11, pancoast12} and \citet{brewer11b}.
The goal of this approach is not to model the physics of the gas in the BLR, but rather to obtain a description of the geometry and kinematics of the gas emission.
The processes at work within the BLR are likely very complex, and an exhaustive BLR model including numerical simulations would be computationally expensive and time consuming.
By using a simple, flexibly parameterized model with a small number of parameters, one can quickly produce emission-line time series and use Markov Chain Monte Carlo methods to put quantitative constraints on the kinematic and geometrical model parameters.
Realistic uncertainties can still be estimated by inflating the error bars on the spectra with a parameter $T$, accounting for the limitations of a simplified model.

The dynamical modeling codes described by \citet[][used in this work]{pancoast14a} and \citet{li13} have so far been applied to 17 AGN \citep{pancoast14b, Pancoast++18, Grier++17, Williams++18, Li++18}.
Each BLR in this sample is best fit with models resembling thick disks that are inclined slightly to the observer, despite there being no preference for this geometry built into the modeling code, and all $M_{\rm BH}$ measurements are consistent with those of other techniques.
The flexibility of the model is apparent in other parameters, such as model kinematics ranging from mostly inflow to mostly outflow.
These applications of dynamical modeling have been limited, however, to a single emission line, \Hb\ $\lambda4861$.
Studies of the higher-ionization lines have not been possible due to the lack of the high-quality UV data required for such modeling.

The applications of the modeling approach have all used the optical continuum as a proxy for the ionizing continuum, as all ground-based reverberation mapping studies must do.
Recent work monitoring continuum emission at a range of wavelengths has shown a measurable lag between the UV fluctuations and the optical continuum fluctuations \citep{agnstorm2, agnstorm3}, raising the question of whether the optical continuum is a suitable proxy for the ionizing continuum.
In the case of black hole mass measurements based on a scale factor $f$, the lag is, to first order, removed in the calibration of $f$ with the $M_{\rm BH}-\sigma_*$ relation.
This is not the case for the dynamical modeling approach, however, and it is unclear how the continuum light curve choice affects the modeling results.

The AGN Space Telescope and Optical Reverberation Mapping (AGN STORM) Project provides a unique data set that can allow us to address some of the modeling assumptions and extend the modeling approach to higher-ionization portions of the BLR.
The AGN STORM Project was anchored by nearly daily observations of the Seyfert 1 galaxy NGC 5548 for six months in 2014 with the {\em Hubble Space Telescope} Cosmic Origins Spectrograph \citep{agnstorm1}.
Concurrent UV and X-ray monitoring was provided by {\em Swift} \citep{agnstorm2}.
Ground-based photometry \citep{agnstorm3} and spectroscopy \citep{agnstorm5} was carried out at a large number of observatories and the UV--optical data were used to study the structure of the accretion disk \citep{agnstorm6}.
The UV spectra revealed both broad and narrow absorption features of unusual strength compared to historical UV observations of NGC 5548 and this required careful modeling of the emission and absorption features \citep{agnstorm8} that will be essential for this paper.
These models were also used to recover velocity--delay maps \citep{agnstorm9} for the strong emission lines that are the subject of this paper.
Much of the analysis of the AGN STORM data has been with the aim of understanding an anomalous period during the middle of the observing campaign when the emission and absorption lines at least partially decoupled from the continuum behavior, the so-called ``BLR holiday'' \citep{agnstorm4,agnstorm7,agnstorm10}.
In this work, we use both the UV and optical continuum light curves to examine the effect of continuum wavelength choice on the modeling results, and we model the BLRs for three emission lines: \Hb, \civ, and \La.

In Section \ref{sect:data}, we provide a brief overview of the data we use for the modeling, and in Section \ref{sect:method}, we summarize the modeling method used.
In Section \ref{sect:results}, we present the modeling results for the \Hb, \civ, and \La\ BLRs, and in Section \ref{sect:jointinference}, we combine the results to make a joint inference on the black hole mass in NGC 5548.
In Section \ref{sect:discussion}, we discuss how the continuum light curve choice affects the modeling results, compare the \Hb\ results to previous modeling, and discuss the similarities and differences of the three line-emitting regions.
Finally, we conclude in Section \ref{sect:summary}.


\section{Data}
\label{sect:data}

\subsection{Continuum light curves}
\label{sect:data_continuum}
We fit models to the data using two separate continuum light curves.
We use a UV light curve to fit models for all three of the emission lines, plus a $V$-band light curve to fit models to the \Hb\ light curve.
Since the UV light curve is a closer proxy to the actual ionizing continuum, we expect this to be the more realistic physical model.
However, the UV is inaccessible to ground-based reverberation mapping campaigns targeting \Hb, and an optical continuum is typically used in its place.
Using both continuum light curves allows us to study the effect this has on modeling results.

The UV continuum light curve is constructed by joining the \emph{HST} 1157.5~\AA~light curve with the \emph{Swift} UVW2 light curve.
Including the \emph{Swift} data allows us to extend the light curve back in time to explore the possibility of longer emission line lags.
Details of the \emph{HST} and \emph{Swift} campaigns can be found in the papers by \citet[][Paper I]{agnstorm1} and \citet[][Paper II]{agnstorm2}, respectively.
To combine the light curves, we scale the \emph{Swift} UVW2 light curve to match the \emph{HST} flux where data overlap in time, and shift the scaled \emph{Swift} light curve by 0.8 days, the time lag between the \emph{Swift} UVW2 and \emph{HST} 1157.5~\AA~light curves as measured by \citet[][Paper III]{agnstorm3}.
The final UV light curve is then the portion of the \emph{Swift} light curve that lies before the start of the \emph{HST} campaign, plus the full \emph{HST} light curve.

The $V$-band light curve data consist of approximately daily observations obtained with several ground-based telescopes between 2013 December and 2014 August.
The details of the optical continuum observing campaign are described by \citet{agnstorm3}.

\subsection{Emission lines}
We model the line-emitting regions producing three lines---\La, \civ, and \Hb.
The raw data for \La\ and \civ\ were obtained using the Cosmic Origins Spectrograph \citep[COS,][]{Green++12} on \emph{HST} from 1 February to 27 July 2014.
Due to the strong absorption features in the UV lines that can influence our modeling results, we use the broad emission line models of \La\ and \civ\ from \citet[][Paper VIII]{agnstorm8}.
The emission lines we use in this paper are the sum of several Gaussian components, namely components 30-38 for \civ\ and components 5-9 for \La.
The uncertainties are then calculated following the prescription of \citet{agnstorm8}.

The $\sim$$15,000$ resolving power of \emph{HST} COS renders modeling the UV lines at full resolution computationally infeasible given our current BLR model.
We therefore bin the \La\ and \civ\ spectra by a factor of 32 in wavelength to reduce this computational load.
Since we are only interested in the larger-scale features of the BLR and emission-line profile, no relevant information is lost in this step.
For \civ\ (\La), we model the spectra from $1500.8 - 1648.6$~\AA\ ($1180.7 - 1278.8$~\AA) in observed wavelength, giving 95 (80) pixels across the binned spectrum.
In LOS velocity, this is -14,000 to 13,900~km/s (-13,600 to 10,100~km/s).

The optical spectroscopic observing campaign is described in detail by \citet[][Paper V]{agnstorm5} and is summarized briefly here.
The \Hb\ spectra were obtained from 2014 January 4 through 2014 July 6 with roughly daily cadence using five telescopes. 
The resulting spectra were decomposed into their individual components to isolate the \Hb\ emission from other emission features in the spectral region.
\citet{agnstorm5} fit models using three templates for \feii, but found the template from \citet{kovacevic10} provided the best fits.
We therefore use this version of the spectral decomposition for this work.
To produce the spectra used in this work, we take the observed spectra and subtract off all modeled components except for the \Hb\ components. 
There are strong \oiii\ residuals at wavelengths longer than $5010$~\AA, so we only model the spectra from $4775.0 - 5008.75$~\AA~in observed wavelength, totaling 188 pixels across the emission line.
In LOS velocity, this is -10,300 to 3,900~km/s.
While this means we do not use the information contained in the spectra redward of $5008.75$~\AA\ to constrain the BLR model, the model still produces a full emission-line profile including the red wing.

\subsection{Anomalous emission line behavior}
As discussed in several of the papers in this series, the broad emission lines appear to stop tracking the continuum light curve part way through the observing campaign \citep{agnstorm4,agnstorm7,agnstorm10}.
Our model of the BLR assumes that the BLR particles respond linearly and instantaneously to all changes in the continuum flux.
Since the anomalous behavior of NGC 5548 is a direct violation of this assumption, we fit our models using only the portion of the spectroscopic campaign in which the BLR appears to be behaving normally.
For this work, we use a cutoff date of ${\rm THJD} = 6743$ (${\rm THJD} = {\rm HJD} - 2,450,000$), as determined for \Hb\ by \citet{agnstorm5}.
The time of de-correlation was measured to be slightly later at ${\rm THJD} = 6766$ for \civ, but for continuity we use the ${\rm THJD} = 6743$ cutoff for all three lines.
In the case of \Hb, we also attempt to model the full spectral time series, but these models fail to converge.


\section{The Geometric and Dynamical Model of the Broad Line Region}
\label{sect:method}

We fit the same BLR model to all three emission lines, allowing us to directly compare the parameters for each line-emitting region.
A full description of the BLR model is given by \citet{pancoast14a}, and a summary is provided here.

\subsection{Geometry}
The BLR is modeled as a distribution of massless point-like particles surrounding a central ionizing source at the origin.
These are not particles meant to represent real BLR gas, but rather a way to represent emission line emissivity in the BLR.
The point particles are assigned radial positions, drawn from a Gamma distribution
\begin{align}
p(r | \alpha, \theta) \propto r^{\alpha - 1} \exp\left(-\frac{r}{\theta}\right)
\end{align}
and shifted from the origin by the Schwarzschild radius $R_s = 2 G M_{\rm BH}/ c^2$ plus a minimum radius $r_{\rm min}$.
To work in units of the mean radius, $\mu$, we perform a change of variables from ($\alpha$, $\theta$, $r_{\rm min}$) to ($\mu$, $\beta$, $F$)
\begin{align}
\mu &= r_{\rm min} + \alpha\theta, \\
\label{eqn:beta}
\beta &= \frac{1}{\sqrt{\alpha}}, \\
F &= \frac{r_{\rm min}}{\mu},
\end{align}
where $\beta$ is the shape parameter and $F$ is the minimum radius ($r_{\rm min}$, typically a few light days) in units of $\mu$.
We assume that the observing campaign is sufficiently long enough to measure time lags throughout the whole BLR, so we truncate the BLR at an outer radius $r_{\rm out} = c\Delta t_{\rm data} / 2$, where $\Delta t_{\rm data}$ is the time between the first continuum light curve model point and the first observed spectrum.
Note that this is not an estimate of the outer edge of BLR emission, and for all cases with campaigns of sufficient duration, the emission trails to near-zero at much smaller radii than $r_{\rm out}$.
The values of $r_{\rm out}$ are reported in Table \ref{tab:results}.

Next, the full plane of particles is inclined relative to the observer's line of sight by an angle $\theta_i$, such that a BLR viewed face-on would have $\theta_i = 0$ deg.
The particles are distributed around this plane with a maximum height parameterized by a half-opening angle $\theta_o$.
The angle above the BLR midplane for an individual particle as seen from the black hole is given by
\begin{align}
\label{eqn:gamma}
\theta = \arccos(\cos \theta_o + (1 - \cos \theta_o) U^\gamma),
\end{align}
where $U$ is drawn from a uniform distribution between 0 and 1 and $\gamma$ is a free parameter between 1 and 5.
In the case of $\gamma = 1$, the point particles are evenly distributed between the central plane and the faces of the disk at $\theta_o$, while for $\gamma = 5$, the particles are clustered at $\theta_o$.

The emission from each individual particle is assigned a weight between 0 and 1 according to 
\begin{align}
\label{eqn:kappa}
W(\phi) = \frac{1}{2} + \kappa\cos(\phi),
\end{align}
where $\phi$ is the angle measured between the observer's line to the origin and the particle's line to the origin, and $\kappa$ is a free parameter between $-0.5$ and $0.5$.
For $\kappa \rightarrow -0.5$, particles preferentially emit back towards the ionizing source, and for $\kappa \rightarrow 0.5$, particles preferentially emit away from the ionizing source.

Additionally, we allow for the presence of an obscuring medium in the plane of the BLR, such as an optically thick accretion disk, that can block line emission from the far side.
The mid-plane can range from transparent to opaque according to the free parameter $\xi$, ranging from 0 (fully opaque) to 1 (fully transparent).
To improve computation time, this is achieved by reflecting a fraction of the particles across the BLR midplane from the far side to the near side.

\subsection{Dynamics}
The wavelength of emission from each particle is determined by the velocity component along the observer's line of sight.
To determine the velocities, we first split the particles into two subsets.
A fraction $f_{\rm ellip}$ are set to have near-circular elliptical orbits around the black hole, with radial and tangential velocities drawn from Gaussian distributions centered on the circular velocity in the $v_r - v_\phi$ plane.
Since the circular velocity depends on the particle position and the black hole mass, $M_{\rm BH}$ enters as a free parameter in this step.

The remaining $1 - f_{\rm ellip}$ particles are assigned to have either inflowing or outflowing trajectories.
In this case, the velocity components are drawn from a Gaussian centered on the radial inflowing or outflowing escape velocity in the $v_r - v_\phi$ plane \citep[see][Figure 2, for an illustration]{pancoast14a}.
Inflow or outflow is determined by the binary parameter $f_{\rm flow}$, where $f_{\rm flow} < 0.5$ indicates inflow and $f_{\rm flow} > 0.5$ indicates outflow.
Additionally, we rotate the velocity components by an angle $\theta_e$ in the $v_r - v_\phi$ plane towards the circular velocity, increasing the fraction of bound orbits as $\theta_e$ increases towards 90 degrees.

We include a contribution from macroturbulent velocities with magnitude
\begin{align}
\label{eqn:sigmaturb}
v_{\rm turb} = \mathcal{N}(0, \sigma_{\rm turb})|v_{\rm circ}|,
\end{align}
where $v_{\rm circ}$ is the circular velocity and $\mathcal{N}(0, \sigma_{\rm turb})$ is the normal distribution with mean $0$ and standard deviation $\sigma_{\rm turb}$, a free parameter.
This value is calculated for each particle and added to its line-of-sight velocity.

Doublet emission lines are accounted for by producing flux shifted in wavelength relative to both doublet rest wavelengths.
Thus, the particles in the \civ\,$\lambda \lambda 1548,1550$ BLR model use both $1548$~\AA\ and $1550$~\AA\ as the reference wavelength.

\subsection{Producing Emission-Line Spectra}
The ionizing source is assumed to be a point source at the origin that emits isotropically and directly follows the AGN continuum light curves described in Section \ref{sect:data_continuum}.
This light propagates out to the BLR particles which instantaneously reprocess the light and convert it into emission line flux seen by the observer.
There is a time-lag between the continuum emission and the line emission determined by the particles' positions, and the wavelength of the light is Doppler shifted from the central emission line wavelength based on the particle's line-of-sight velocity.
In the case of \civ, both components of the doublet emission line are included.

Since the BLR particles can lie at arbitrary distances from the central ionizing source, we need a way to calculate the continuum flux at arbitrary times.
We use Gaussian processes as a means of flexibly interpolating between points in the observed continuum light curve as well as extending the light curve to times before or after the start of the campaign to explore the possibility of longer lags.
The Gaussian process model parameters are included in our parameter exploration which allows us to include the continuum interpolation uncertainty in our inference of the other BLR model parameters.

\subsection{Exploring the Model Parameter Space}
For each set of model parameters, we use 4000 BLR test particles to produce an emission-line time series with times corresponding to the actual epochs of observation.
We can compare the observed spectra with the model spectra using a Gaussian likelihood function and adjust the model parameters accordingly.
To explore the BLR and continuum model parameter space, we use the diffusive nested sampling code {\sc DNest4} \citep{dnest4}.
Diffusive nested sampling is a Markov Chain Monte Carlo method based on Nested Sampling that is able to efficiently explore high-dimensional and complex parameter spaces.

{\sc DNest4} allows us to do further analysis in post-processing through the introduction of a temperature $T$, which softens the likelihood function by dividing the log of the likelihood by $T$.
The temperature in this case is not a physical temperature, but rather a parameter commonly used in optimization algorithms such as simulated annealing \citep{Kirkpatrick++83}.
In the case of a Gaussian likelihood function, this is equivalent to multiplying the uncertainties on the observed spectra by $\sqrt{T}$.
This factor can account for under-estimated uncertainties on the spectra or the inability of the simplified model to accurately fit the complexities of the real data.

\newcommand{\resultstablecomment}{Median and 68\% confidence intervals for the main BLR model parameters. Note that $r_{\rm out}$ is a fixed parameter, so we do not include uncertainties, and we also include the temperature $T$ used in post-processing.}
\begin{deluxetable*}{llcccc}
\tablecaption{BLR Model Parameter Values}
\tablehead{ 
\colhead{Parameter} &
\colhead{Brief Description} &
\colhead{\La} &
\colhead{\civ} &
\colhead{\Hb\ vs. UV} &
\colhead{\Hb\ vs. $V$-band} 
}
\startdata
$\log_{10}(M_{\rm bh}/M_{\odot})$ & Black hole mass & $7.38^{+0.54}_{-0.41}$ & $7.58^{+0.33}_{-0.21}$ & $7.72^{+0.20}_{-0.18}$ & $7.54^{+0.34}_{-0.24}$\\
$r_{\rm mean}$ (${\rm light~days}$) & Mean line emission radius & $12.3^{+5.0}_{-4.4}$ & $11.2^{+2.7}_{-2.3}$ & $12.2^{+6.3}_{-5.1}$ & $8.0^{+4.3}_{-2.6}$\\
$r_{\rm median}$ (${\rm light~days}$) & Median line emission radius & $4.0^{+2.4}_{-1.7}$ & $3.5^{+1.3}_{-0.8}$ & $9.1^{+5.2}_{-3.8}$ & $6.1^{+3.7}_{-2.1}$\\
$r_{\rm min}$ (${\rm light~days}$) & Minimum line emission radius & $1.08^{+0.80}_{-0.49}$ & $1.17^{+0.42}_{-0.29}$ & $3.85^{+1.99}_{-2.14}$ & $2.38^{+1.96}_{-0.99}$\\
$\sigma_r$ (${\rm light~days}$) & Radial width of line emission & $23.3^{+15.3}_{-9.6}$ & $20.1^{+6.8}_{-4.8}$ & $11.7^{+11.7}_{-5.9}$ & $6.8^{+9.1}_{-2.4}$\\
$\tau_{\rm mean}$ (${\rm days}$) & Mean lag in observer frame & $11.6^{+4.5}_{-4.7}$ & $11.3^{+2.4}_{-2.2}$ & $9.9^{+5.1}_{-3.8}$ & $7.0^{+3.2}_{-2.3}$\\
$\tau_{\rm median}$ (${\rm days}$) & Median lag in observer frame & $3.6^{+1.9}_{-1.7}$ & $3.3^{+1.1}_{-0.7}$ & $7.1^{+3.1}_{-2.7}$ & $4.8^{+2.3}_{-1.7}$\\
$\beta$ & Shape parameter of radial distribution (Eqn. \ref{eqn:beta}) & $1.86^{+0.10}_{-0.14}$ & $1.89^{+0.07}_{-0.15}$ & $1.17^{+0.23}_{-0.24}$ & $1.12^{+0.22}_{-0.18}$\\
$\theta_o$ (${\rm degrees}$) & Half-opening angle & $31.9^{+20.5}_{-12.2}$ & $30.9^{+8.0}_{-7.9}$ & $35.8^{+13.8}_{-7.4}$ & $38.6^{+14.0}_{-13.5}$\\
$\theta_i$ (${\rm degrees}$) & Inclination angle & $23.7^{+23.6}_{-9.0}$ & $28.3^{+8.1}_{-9.2}$ & $46.1^{+13.4}_{-9.0}$ & $47.3^{+13.0}_{-15.8}$\\
$\kappa$ & Cosine illumination function parameter (Eqn. \ref{eqn:kappa}) & $-0.23^{+0.52}_{-0.24}$ & $-0.42^{+0.12}_{-0.06}$ & $0.00^{+0.10}_{-0.08}$ & $-0.01^{+0.09}_{-0.07}$\\
$\gamma$ & Disk face concentration parameter (Eqn. \ref{eqn:gamma}) & $3.5^{+1.1}_{-1.5}$ & $4.1^{+0.7}_{-1.3}$ & $3.4^{+1.1}_{-1.4}$ & $3.0^{+1.3}_{-1.3}$\\
$\xi$ & Mid-plane transparency & $0.33^{+0.45}_{-0.25}$ & $0.44^{+0.31}_{-0.27}$ & $0.20^{+0.17}_{-0.15}$ & $0.17^{+0.21}_{-0.12}$\\
$f_{\rm ellip}$ & Elliptical orbit fraction & $0.20^{+0.16}_{-0.13}$ & $0.23^{+0.17}_{-0.15}$ & $0.29^{+0.18}_{-0.18}$ & $0.29^{+0.18}_{-0.20}$\\
$f_{\rm flow}$ & Inflow/outflow flag & $0.60^{+0.29}_{-0.40}$ & $0.41^{+0.40}_{-0.27}$ & $0.74^{+0.19}_{-0.19}$ & $0.73^{+0.18}_{-0.17}$\\
$\theta_e$ (${\rm degrees}$) & Angle in $v_r - v_\phi$ plane & $29^{+20}_{-19}$ & $26^{+15}_{-17}$ & $39^{+19}_{-15}$ & $42^{+16}_{-21}$\\
$\sigma_{\rm turb}$ & Turbulence (Eqn. \ref{eqn:sigmaturb}) & $0.018^{+0.049}_{-0.016}$ & $0.008^{+0.033}_{-0.006}$ & $0.022^{+0.055}_{-0.019}$ & $0.029^{+0.038}_{-0.026}$\\
$r_{\rm out}$ $({\rm light~days})$ & Outer line emission radius (fixed parameter) & $145$ & $145$ & $81$ & $80$\\
$T$ & Temperature (statistical) & $5000$ & $500$ & $300$ & $200$\\
\enddata
\tablecomments{\resultstablecomment
\label{tab:results}}
\end{deluxetable*}

The value of $T$ is determined by examining the sample distributions at increasing levels of likelihood and choosing the largest $T$ for which the distributions remain smooth and do not contain several local minima. 
The choices of $T$ for each run are listed in Table \ref{tab:results}.
In the cases of \La\ and \civ, we required very large temperatures due to the inability of the simple model to fit the level of detail present in the high-SNR \emph{HST} data.

Convergence of the modeling runs was determined by ensuring that the parameter distributions for the second half of each run matched the parameter distribution for the first half of the run.


\section{Results}
\label{sect:results}
In this section, we describe the results of fitting our BLR model to the data.
For each emission line, we give the posterior probability density functions (PDFs) for the model parameters and use these to draw inferences on the structural and kinematic properties of the BLRs. 
From the posterior samples, we show one possible geometric structure of the BLR gas emission, selected to have parameter values closest to the median inferred values.
We also show the transfer function, $\Psi(\lambda,\tau)$, which describes how continuum ($C$) fluctuations are mapped to emission line ($L$) fluctuations as a function of wavelength and time-delay:
\begin{align}
L(\lambda,t) = \int \Psi(\lambda, \tau) C(t - \tau) d\tau.
\end{align}
The functions shown are calculated by producing transfer functions for 30 random models from the posterior and calculating the median value in each wavelength-delay bin.
Table \ref{tab:results} lists the inferred model parameters for each line-emitting region.

\subsection{\Hb}\label{sect:hbeta}
Multi-wavelength monitoring campaigns have shown that longer continuum wavelengths tend to lag behind shorter wavelengths \citep[e.g., ][]{agnstorm2, Edelson++17, agnstorm3, Fausnaugh++18}, indicating that the UV is a closer proxy to the ionizing continuum than the $V$ band.
Additionally, the shorter-wavelength continuum variations show more short-timescale structure than longer wavelengths.
Since the emission lines respond to the short-timescale ionizing continuum variations, one could observe higher-frequency emission-line variability than is present in the smoothed $V$-band continuum light curve.
Complicating matters even further, recent studies have shown that diffuse continuum emission arising in the BLR gas can be strong enough to significantly enhance continuum lags, especially at optical wavelengths \citep{Korista+01, Cackett++18, Lawther++18, Korista+19}.

\begin{figure}[hb!]
\begin{center}
\includegraphics[width=3.45in]{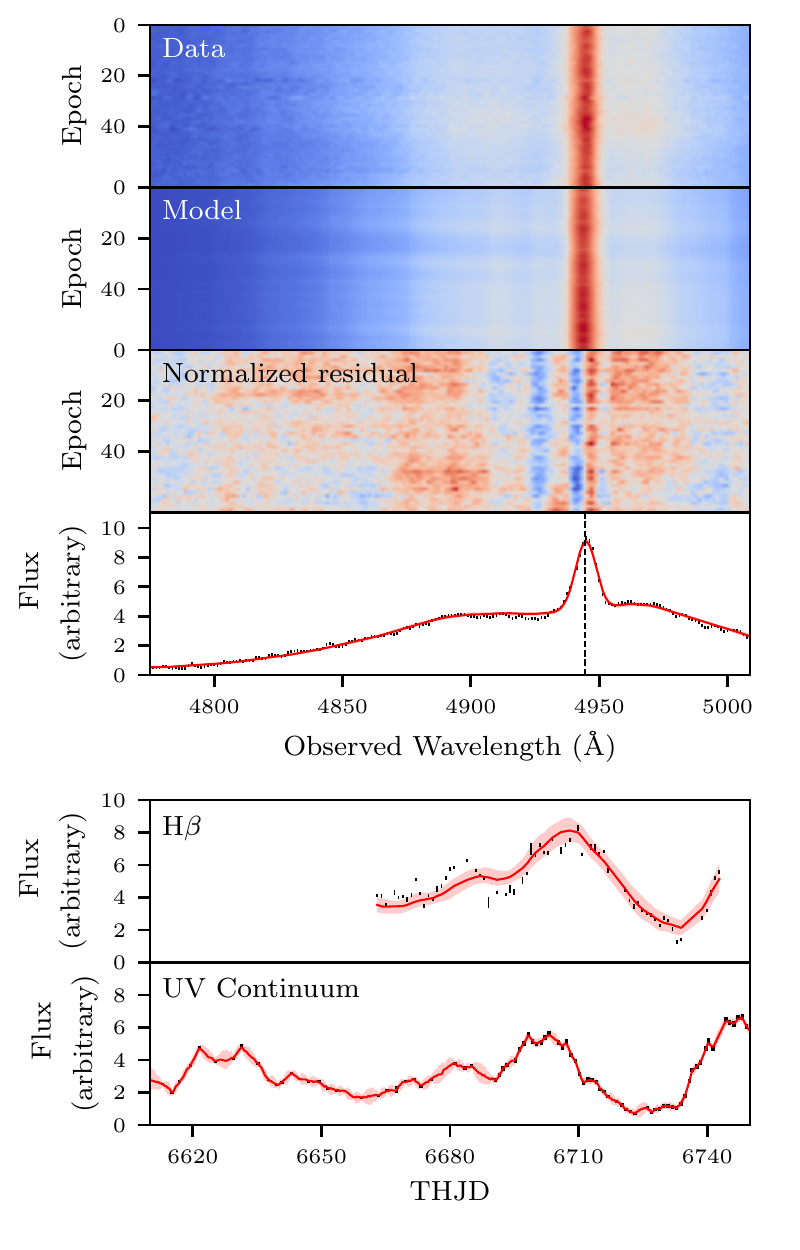}
\caption{Numbered 1-6 from top to bottom, \textit{Panels 1-3}: The observed \Hb\ emission-line profile by observation epoch, the profiles produced by one possible BLR model, and the normalized residual ($[{\rm Data} - {\rm Model}]/{\rm Data~uncertainty}$). \textit{Panel 4}: The observed \Hb\ profile of the tenth epoch (black) and the emission-line profile produced by the model shown in Panel 2 (red). The vertical dashed line shows the emission line center in the observed frame. \textit{Panel 5}: Time series (${\rm THJD} = {\rm HJD} - 2,450,000$) of the integrated \Hb\ emission line data (black) and the integrated \Hb\ model shown in Panel 2 (red). \textit{Panel 6}: The same as Panel 5, but with the continuum flux rather than integrated \Hb\ flux. In Panels 4-6, the light red band shows the 1$\sigma$ scatter of all models in the posterior sample.
\label{fig:display_hb_uv}}
\end{center}
\end{figure}

\begin{figure}[hb!]
\begin{center}
\includegraphics[width=3.3in]{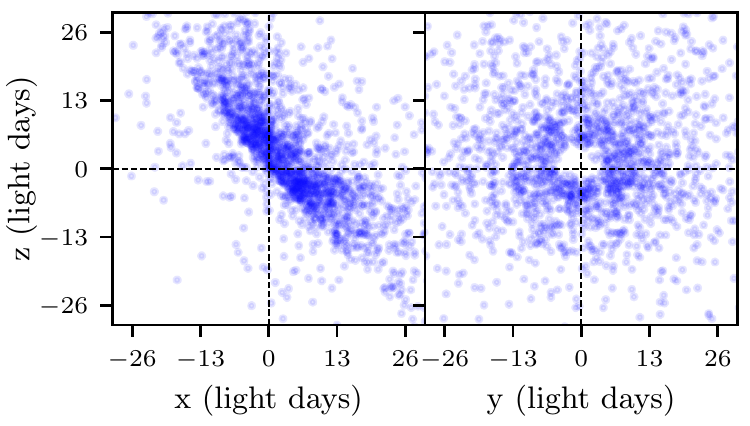} 
\caption{
Possible geometry for the \Hb-emitting BLR, when modeled using the UV light curve.
The left-hand panel shows an edge-on view with the observer on the positive $x$-axis, and the right-hand panel shows a face-on view of the BLR, as seen by the observer.
The size of the circles represents the relative amount of emission from the particles, as seen by the observer.
This value is determined by the particle's position and the parameter $\kappa$ (Equation \ref{eqn:kappa}).
Note that few particles are shown in the bottom-left portion of the left-hand panel due to how the code handles an opaque mid-plane.
\label{fig:geo_hb_uv}}
\end{center}
\end{figure}

When the $V$ band is used, these combined effects can lead to shorter \Hb-optical lags and may result in $M_{\rm BH}$ underestimates if not accounted for.
However, since the UV is not available for ground-based reverberation mapping campaigns, the $V$ band is very often used as a proxy for the ionizing continuum.
Since both light curves are available in the AGN STORM data set, we have a unique opportunity to compare the modeling results using each continuum light curve.
We run our modeling code with the \Hb\ emission line data using both the UV and $V$-band light curves as the driving continuum to study potential systematics introduced by the choice of continuum wavelength.

\subsubsection{\Hb\ vs. UV light curve}

For the first \Hb\ modeling tests, we use the \emph{HST} 1157.5~\AA\ plus \emph{Swift} UVW2 light curve as the driving continuum.
The data require a temperature of $T=300$, equivalent to increasing the spectral uncertainties by a factor of $\sqrt{300} = 17.3$.
As shown in Figure \ref{fig:display_hb_uv}, our model fits the rough shape of the emission line light curve, but there is clear structure in the residuals near the line peak.
Additionally, there is a small trough in the emission line data at wavelengths just short of the line peak that the models are unable to reproduce. 
Looking at the integrated \Hb\ flux light curve, we see that the models can reproduce the general structure of the variations, but the full amplitude of variations is not perfectly matched.
In particular, the fluctuations in the first half of the \Hb\ light curve are larger than those predicted by the models, while the same models are able to reproduce the larger-scale rise and fall in the second half of the light curve.

\begin{figure*}[ht!]
\begin{center}
\includegraphics[width=7in]{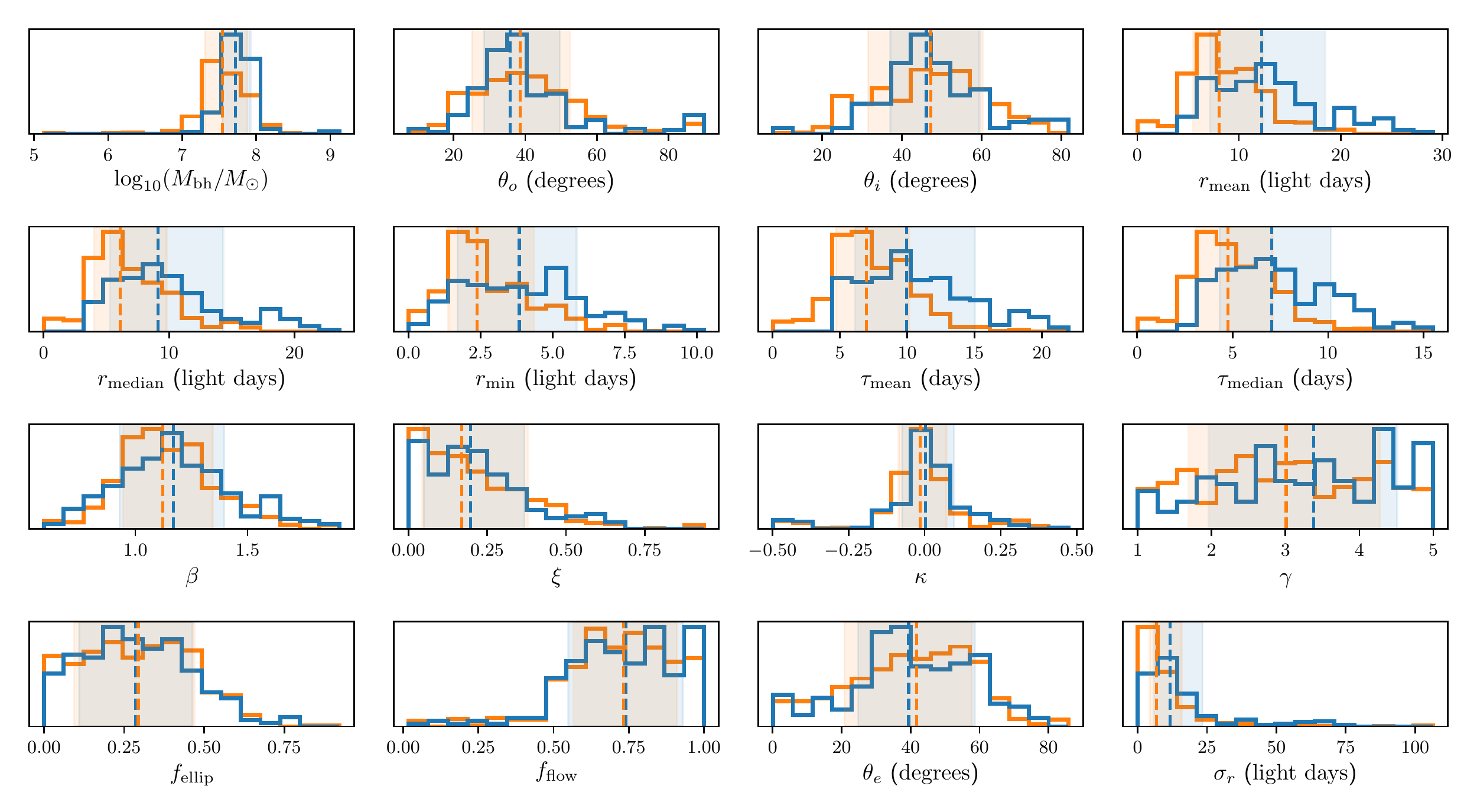}
\caption{Comparison of the posterior PDFs for the BLR model parameters obtained when using the UV (blue) and $V$ band (orange) as the continuum light curve driving the \Hb\ variations.
The vertical dashed lines show the median parameter values, and the shaded regions show the 68\% confidence intervals.
\label{fig:posterior_hb_comparison}}
\end{center}
\end{figure*}

\begin{figure*}[h!]
\begin{center}
\includegraphics[width=3in]{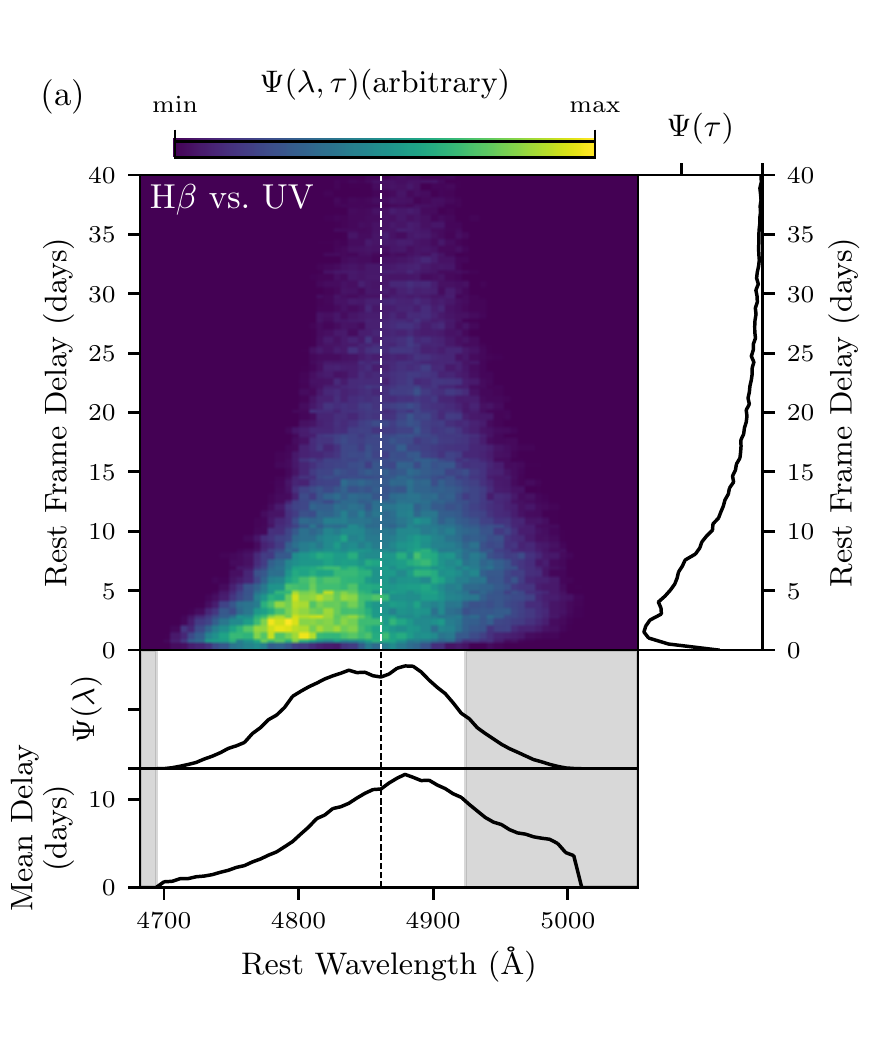} 
\includegraphics[width=3in]{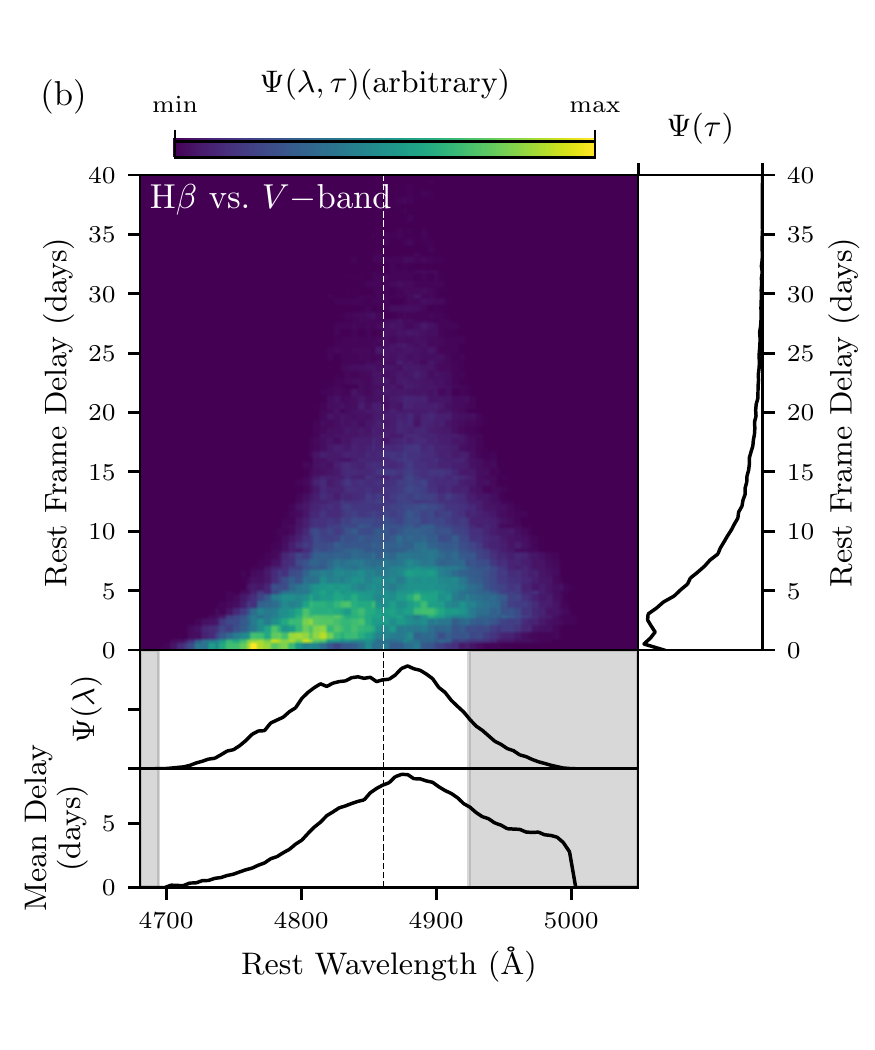} 
\includegraphics[width=3in]{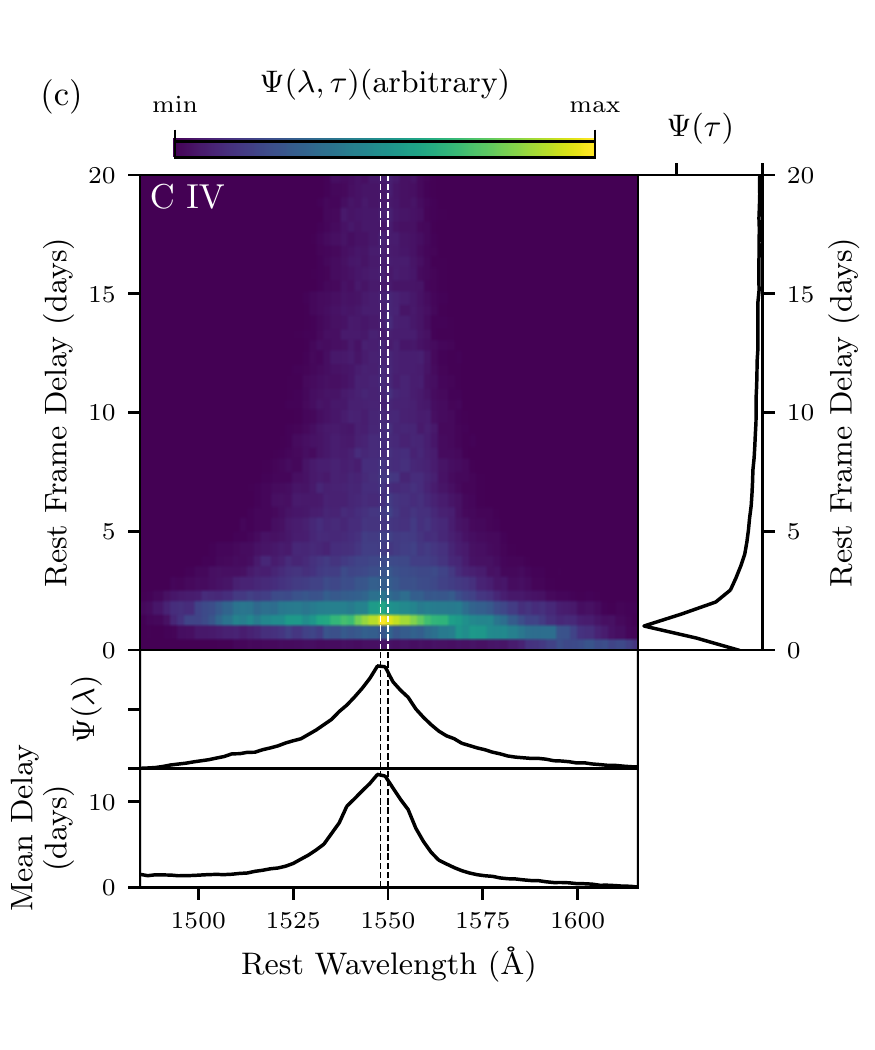} 
\includegraphics[width=3in]{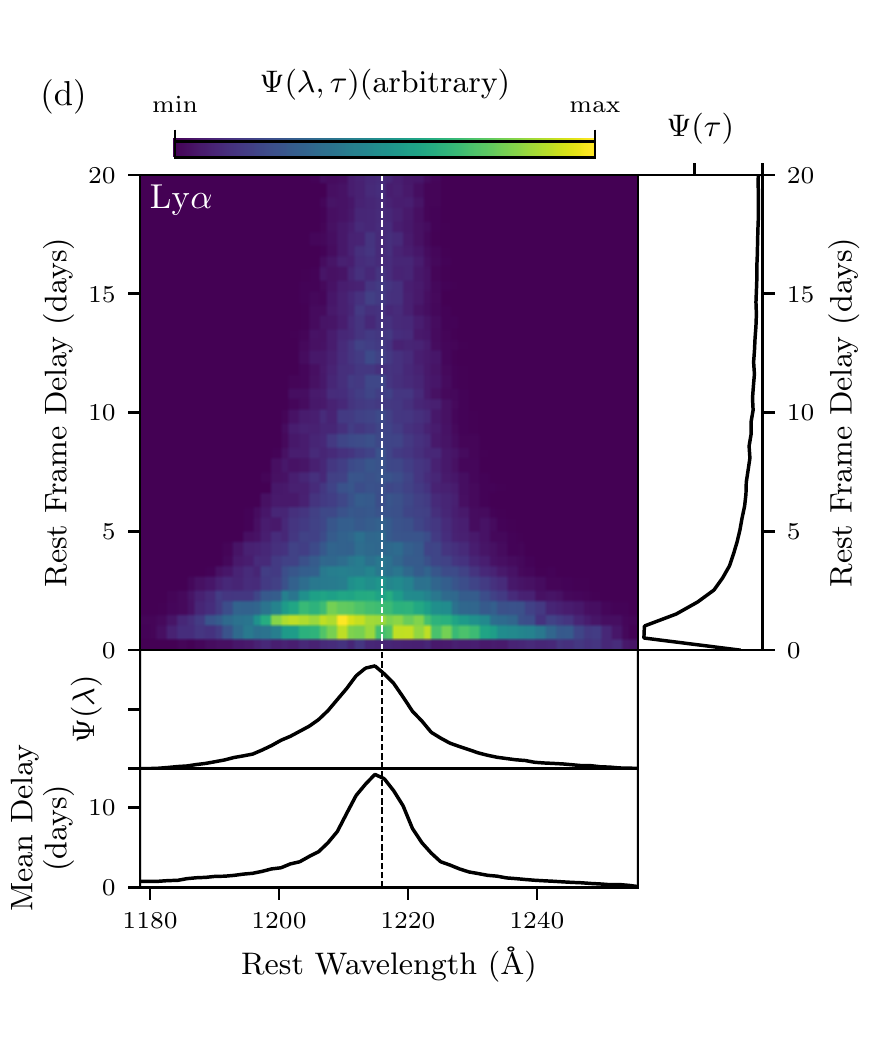} 
\caption{
Median transfer functions for each BLR, calculated by producing transfer functions for 30 random models from the posterior and calculating the median value in each wavelength-delay bin.
The bottom panels show the lag-integrated transfer function, $\Psi(\lambda)$, and the mean rest frame lag as a function of wavelength.
The right-hand panel shows the velocity-integrated response, $\Psi(\tau)$, as a function of rest frame lag.
The greyed out regions indicate the wavelength range that was not modeled for \Hb, and vertical dashed lines show the emission line center.
\label{fig:transfer}}
\end{center}
\end{figure*}

Geometrically, we find a BLR that has a thick disk structure that is highly inclined relative to the observer (Figure \ref{fig:geo_hb_uv}).
The opening angle posterior PDF has a primary peak at $35$ degrees and a small secondary peak near 90 degrees (Figure \ref{fig:posterior_hb_comparison}, blue lines).
Similarly, the inclination angle posterior PDF has a primary peak at $45$ degrees and a small secondary rise towards 80 degrees. 
Simply taking the median and 68\% confidence intervals for these parameters gives $\theta_o = 35.8^{+13.8}_{-7.4}$ degrees and $\theta_i = 46.1^{+13.4}_{-9.0}$ degrees.

The median radius of the BLR is $r_{\rm median} = 9.1^{+5.2}_{-3.8}$ light days with an inner minimum radius of $r_{\rm min} = 3.9^{+2.0}_{-2.1}$ light days.
The radial width of the BLR is $\sigma_r = 11.7^{+11.7}_{-5.9}$ light days, and the radial distribution of BLR particles is close to exponential with $\beta = 1.17^{+0.23}_{-0.24}$.
The relative distribution of particles within the disk (either uniformly distributed or concentrated near the opening angle) is not constrained ($\gamma = 3.4^{+1.1}_{-1.4}$).
We find a preference for isotropic emission from all BLR particles, rather than emission back towards or away from the ionizing source ($\kappa = 0.00^{+0.10}_{-0.08}$).
In previous modeling of the \Hb\ BLR in other AGN \citep{pancoast14b, Pancoast++18, Grier++17, Williams++18}, nearly every object in which $\kappa$ is well determined has $\kappa < 0$ at the 1$\sigma$ level or greater.
This is also the result that is predicted from photoionization models, so we discuss the value from this work further at the end of the section.
Finally, models with an opaque midplane are preferred over those without, with $\xi = 0.20^{+0.17}_{-0.15}$.

Kinematically, the data prefer models in which a third of the BLR particles are on elliptical orbits ($f_{\rm ellip} = 0.29^{+0.18}_{-0.18}$).
The remaining particles are mostly outflowing, with $f_{\rm flow} = 0.74^{+0.19}_{-0.19}$, although some of these may still be on bound, highly elliptical orbits, with $\theta_e = 39^{+19}_{-15}$ degrees.
We find little contribution from macroturbulent velocities, with $\sigma_{\rm turb} = 0.022^{+0.055}_{-0.019}$).
Finally, we measure the black hole mass in this model to be $\log_{10}(M_{\rm BH}/M_\odot) = 7.72^{+0.20}_{-0.18}$.

The \Hb~vs.~UV lag one would measure from the models is $\tau_{\rm median} = 7.1^{+3.1}_{-2.7}$ days.
This agrees with the \citet{agnstorm5} measurements of $\tau_{\rm cen,T1} = 7.62^{+0.49}_{-0.49}$ days from cross-correlation and $\tau_{\rm JAVELIN, T1} = 6.91^{+0.64}_{-0.63}$ days from {\sc JAVELIN} \citep{zu11}.
Both of these measurements used the $F_\lambda(1158~\rm{\AA})$ light curve as the driving continuum and the \Hb\ spectra up to ${\rm THJD} = 6743$, the same dates used to fit our models.
To measure a black hole mass, \citep{agnstorm5} use the cross-correlation lag between \Hb\ and the 5100~\AA\ continuum, and calculate $M_{\rm BH}/10^7 M_\odot = 7.53^{+1.96}_{-1.99}$ ($\log_{10}[M_{\rm BH}/M_\odot] = 7.88^{+0.10}_{-0.13}$), which is consistent with our measurement.

\begin{figure}[hb!]
\begin{center}
\includegraphics[width=3.45in]{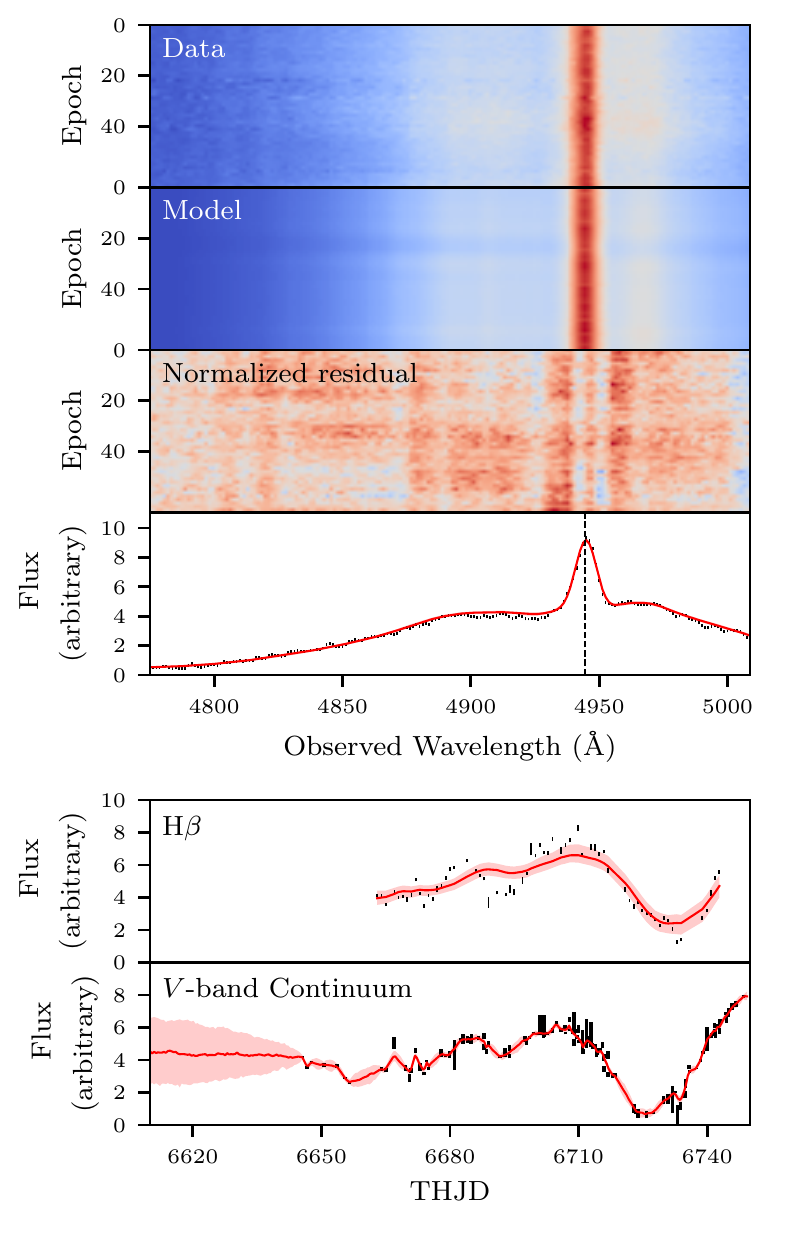}
\caption{Same as Figure \ref{fig:display_hb_uv}, but for the \Hb\ models using the $V$ band as the driving continuum.
The large scatter in modeled continuum light curves before ${\rm THJD} \sim 6650$ is due to extrapolation to times before the monitoring campaign started.
\label{fig:display_hb_vband}}
\end{center}
\end{figure}

\citet{agnstorm9} find velocity-delay maps that interpreted as indicating a BLR with inclination angle $i = 45$ degrees, a 20 light day outer radius with most response between 5 and 15 days, and black hole mass $M_{\rm BH} = 7\times 10^7 M_\odot$ [$\log_{10}(M_{\rm BH}/M_\odot) = 7.8$].
Our black hole mass and inclination angle measurements agree with these values, but we do find models with BLR emission extending to radii greater than 20 light days.
We remind the reader that $r_{\rm out}$ in our model is a fixed parameter determined by the campaign duration and should not be interpreted as a measurement of the BLR outer radius.

The transfer function produced by our model (Figure \ref{fig:transfer}, a) shows that the emission is enclosed within a virial envelope, similar to the maps of \citet{agnstorm9}.
There is a slight angle to the transfer function, showing more emission at short lags and bluer wavelengths, which can be interpreted as an outflow. 
This agrees with the $f_{\rm ellip}$ and $f_{\rm flow}$ values in the model.
Compared with the velocity-resolved measurements of \citet[][Figure 10]{agnstorm5}, our plot of the mean delay is noticeably lacking the distinct `M' shape with short lags at the core of the emission line.
One way to achieve such a shape is if the far side of the BLR does not respond to the continuum, possibly due to an obscurer.
Our simple model is unable to produce such an asymmetric effect, so it is possible that the $\kappa$ parameter was pushed to greater values in order to dampen the response of the far side. 

\subsubsection{\Hb\ vs. $V$-band light curve}
For the second \Hb\ modeling tests, we use the $V$-band light curve as the driving continuum, with the same \Hb\ spectra up until the \citet{agnstorm5} cutoff. 
We use a temperature $T=200$, corresponding to an increase in spectral uncertainties of a factor $\sqrt{200} = 14.1$. 
Similar to the \Hb\ vs. UV models, the \Hb\ vs. $V$ band models are able to reproduce the large-scale shape of the emission-line profile, but they are unable to fit the smaller-scale wiggles (Figure \ref{fig:display_hb_vband}).
Again, the amplitude of fluctuations in the \Hb\ light curve is not fully reproduced in the $V$-band-driven models, although the general structure is still well captured.
In general, the $V$-band-driven models produce integrated emission line light curves that are smoothed compared to the UV-driven counterparts.

\begin{figure}[hb!]
\begin{center}
\includegraphics[width=3.3in]{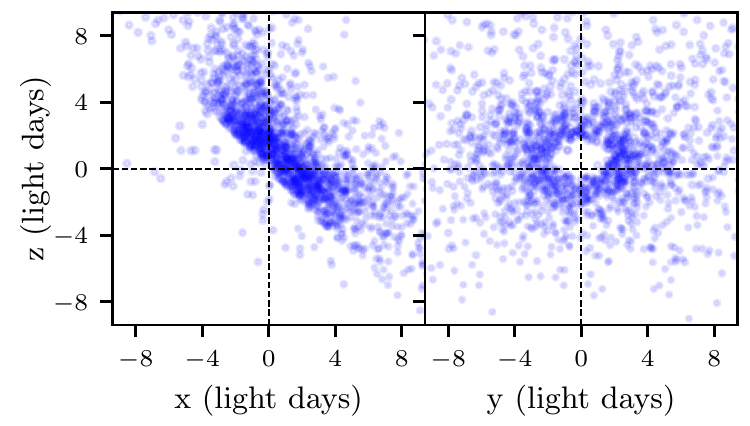}
\caption{
Same as Figure \ref{fig:geo_hb_uv}, but for the \Hb-emitting BLR modeled using the $V$-band light curve as the driving continuum.
\label{fig:geo_hb_vband}}
\end{center}
\end{figure}

Geometrically, models with an inclined thick disk structure are preferred, with $\theta_o = 38.6^{+14.0}_{-13.5}$ degrees and $\theta_i = 47.3^{+13.0}_{-15.8}$ degrees (Figure \ref{fig:geo_hb_vband}).
The median radius is $r_{\rm median} = 6.1^{+3.7}_{-2.1}$ light days, the minimum radius is $r_{\rm min} = 2.4^{+2.0}_{-1.0}$ light days, and the radial width is $\sigma_r = 6.8^{+9.1}_{-2.4}$ light days.
The radial distribution is close to exponential with $\beta = 1.12^{+0.22}_{-0.18}$ and the distribution of particles within the disk is not constrained ($\gamma = 3.0^{+1.3}_{-1.3}$).
The BLR particles emit isotropically ($\kappa = -0.01^{+0.09}_{-0.07}$), and there is a preference for an opaque midplane ($\xi = 0.17^{+0.21}_{-0.12}$). 

\begin{figure}[hb!]
\begin{center}
\includegraphics[width=3.45in]{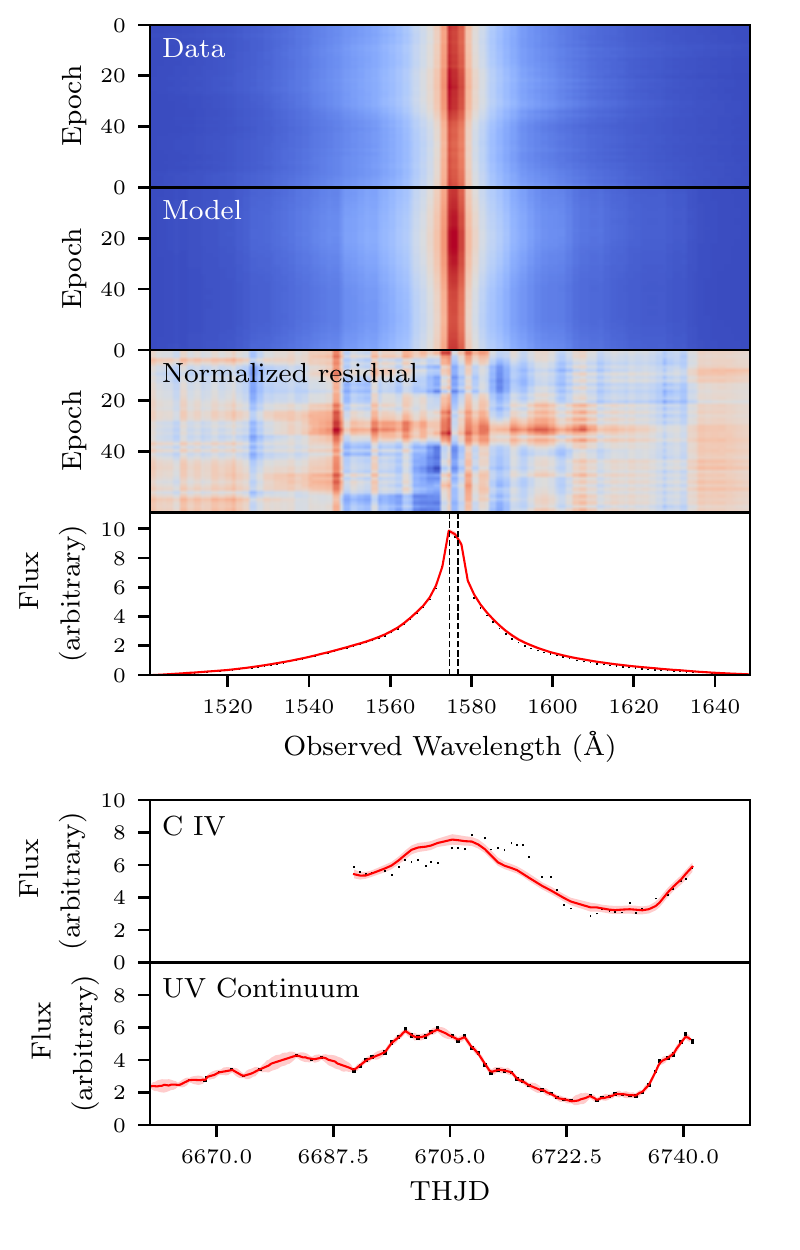}
\caption{Same as Figure \ref{fig:display_hb_uv}, but for the \civ\ BLR models.
\label{fig:display_civ}}
\end{center}
\end{figure}

Dynamically, models with roughly a third of the particles on elliptical motions are preferred ($f_{\rm ellip} = 0.29^{+0.18}_{-0.20}$), and the remaining particles are outflowing $f_{\rm flow} = 0.73^{+0.18}_{-0.17}$, although many may be on highly elliptical bound orbits ($\theta_e = 42^{+16}_{-21}$ degrees).
There is little contribution from macroturbulent velocities, with $\sigma_{\rm turb} = 0.029^{+0.038}_{-0.026}$.
The black hole mass in this model is $\log_{10}(M_{\rm BH}/M_\odot) = 7.54^{+0.34}_{-0.24}$.
The transfer function for this model is very similar to those of the models that use the UV light curve as the driving continuum, but the preference for outflow is slightly more pronounced.

The emission line lag one would measure from the models is $\tau_{\rm median} = 4.8^{+2.3}_{-1.7}$ days.
Within the uncertainties, this agrees with the cross-correlation and JAVELIN measurements of $\tau_{\rm cen,T1} = 3.82^{+0.57}_{-0.47}$ and $\tau_{\rm JAVELIN, T1} = 4.89^{+0.66}_{-0.71}$ days from \citet{agnstorm5}.
Our black hole mass is formally consistent with their measurement of $\log_{10}(M_{\rm BH}/M_\odot) = 7.88^{+0.10}_{-0.13}$, but slightly smaller for the reason described below.

If $M_{\rm BH, UV}$ and $M_{{\rm BH}, V}$ are the masses measured using the UV and $V$-band continua, respectively, we expect to find $M_{\rm BH, UV}/M_{{\rm BH}, V} = \tau_{\rm UV}/\tau_V$.
Since the lag between the UV and $V$-band continua is $\tau_{{\rm UV}-V} = \tau_{\rm UV} - \tau_{V}$, we can write
\begin{multline}
\log_{10}\left(\frac{M_{\rm BH, UV}}{M_\odot}\right) - \log_{10}\left(\frac{M_{{\rm BH}, V}}{M_\odot}\right) \\
= \log_{10}\left(1 + \frac{\tau_{{\rm UV}-V}}{\tau_{V}}\right)
\end{multline}
Using $\tau_{{\rm UV}-V} = 1.86\pm 0.08$ days from \citet{agnstorm3} and $\tau_{V} = \tau_{{\rm median}, V}$, we expect a difference in $\log_{10}(M_{\rm BH}/M_\odot)$ measurements $0.14^{+0.07}_{-0.05}$ solely due to the UV-optical continuum lag.
Our measurements are consistent with this difference.

\subsection{\civ\ (vs. UV light curve)} \label{sect:civ}
The \civ\ emission line has many absorption features that can affect the modeling results.
We therefore use the models from Paper VIII of this series \citep{agnstorm8}, using the components corresponding to the \civ\ emission line.
Due to the high spectral resolution of the data, we also bin the emission line spectra by a factor of 32 in wavelength.
This decreases the run-time of the modeling code not only by reducing the number of data points, but also by reducing the number of BLR test particles that would be required to fit such high resolution data.
We use a temperature of $T=500$, which is equivalent to increasing the uncertainties by a factor of $\sqrt{500} = 22.4$. 

\begin{figure}[hb!]
\begin{center}
\includegraphics[width=3.3in]{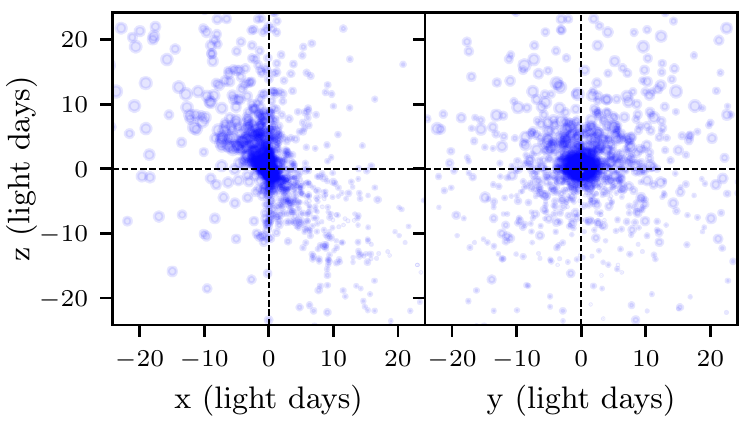} 
\caption{
Same as Figure \ref{fig:geo_hb_uv}, but for the \civ-emitting BLR.
\label{fig:geo_civ}}
\end{center}
\end{figure}

\begin{figure*}[ht!]
\begin{center}
\includegraphics[width=7in]{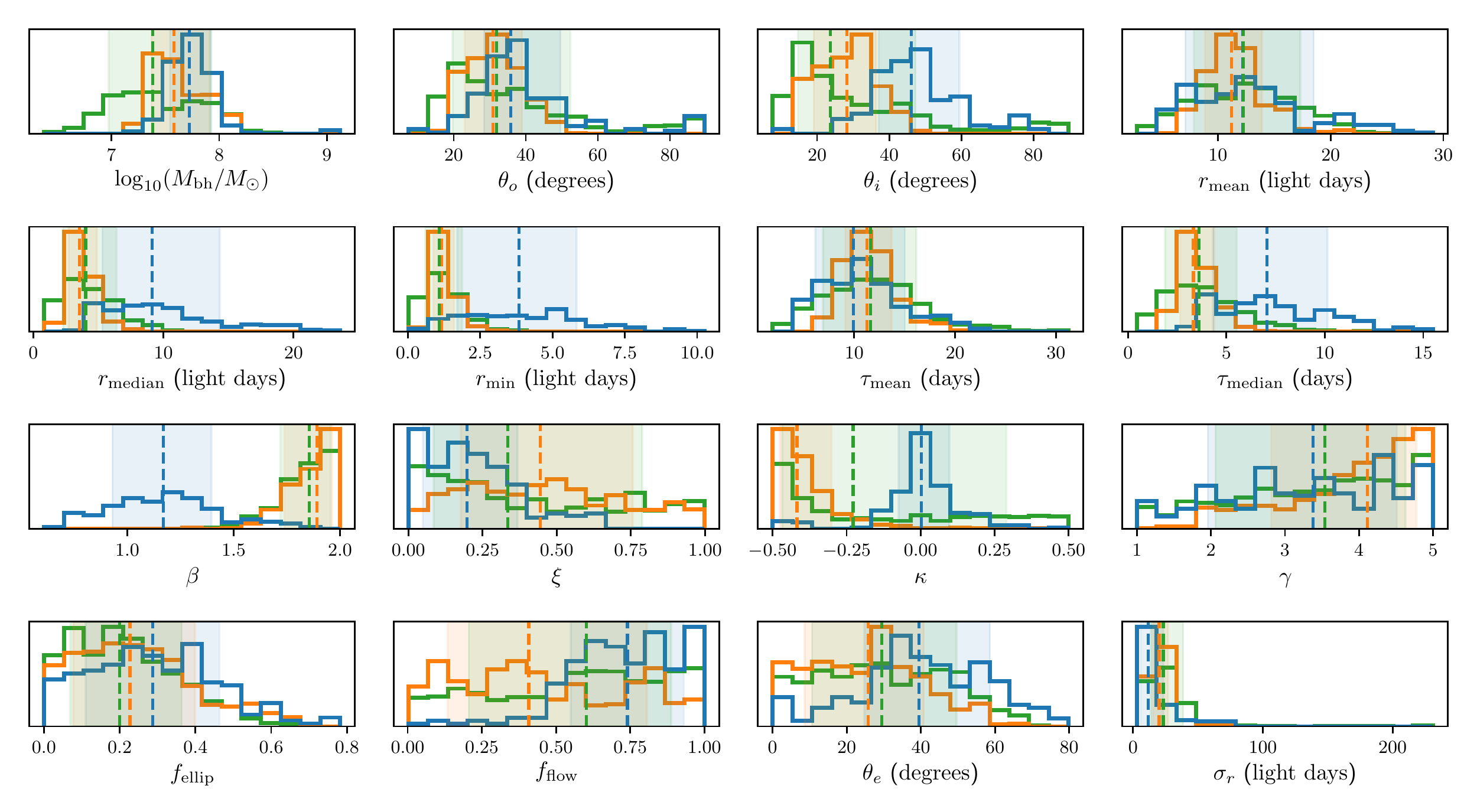}  
\caption{Comparison of the posterior PDFs for the parameters of the \Hb\ (blue), \civ\ (orange), and \La\ (green) BLR models, all using the UV light curve as the driving continuum.
The vertical dashed lines show the median parameter values, and the shaded regions show the 68\% confidence intervals.
\label{fig:linecomparison}}
\end{center}
\end{figure*}

Figure \ref{fig:display_civ} shows the model fits to the \civ\ emission line data.
Note that while the emission line appears to be single-peaked in the figure due to the binning, both peaks are accounted for in the modeling code.
Since the UV emission line light curves are shorter than the ground-based optical emission line light curves, there are fewer features allowing the code to determine the time-lag and hence the radius of the BLR.
The one strong up-and-down fluctuation in the \civ\ light curve is well captured by our model.

Geometrically, the \civ\ BLR has a thick disk structure ($\theta_o = 30.9^{+8.0}_{-7.9}$ degrees, Figure \ref{fig:geo_civ}) that is inclined relative to the observer's line-of-sight ($\theta_i = 28.3^{+8.1}_{-9.2}$ degrees), similar to the results for \Hb\ (Figure \ref{fig:linecomparison}).
The radial distribution, however, has a shape parameter of $\beta = 1.89^{+0.07}_{-0.15}$, indicating a very steep drop-off in the density of BLR emission close to $r_{\rm min}$. 
The median radius of the BLR is $r_{\rm median} = 3.5^{+1.3}_{-0.8}$ light days with an inner minimum radius of $r_{\rm min} = 1.17^{+0.42}_{-0.29}$ light days.
Formally, the standard deviation of the radial distribution of particles is $\sigma_r = 20.1^{+6.8}_{-4.8}$ light days, although this is likely biased high due to the long tails of the distribution.
There is a slight preference for the particles to be concentrated near the opening angle, but this parameter is not well determined ($\gamma = 4.1^{+0.7}_{-1.3}$).
There is a strong preference for emission back towards the ionizing source with $\kappa = -0.42^{+0.12}_{-0.06}$, and there is no preference for an opaque or transparent midplane ($\xi = 0.44^{+0.31}_{-0.27}$).

The data prefer models in which roughly a quarter of the BLR particles are on elliptical orbits ($f_{\rm ellip} = 0.23^{+0.17}_{-0.15}$).
Perhaps surprisingly, \civ\ shows the weakest evidence for outflow, with $f_{\rm flow} = 0.41^{+0.40}_{-0.27}$.
This can be seen in the transfer functions in which there is a weak preference for inflow, with shorter responses at longer wavelengths.
There is little contribution from macroturbulent velocities, with $\sigma_{\rm turb} = 0.008^{+0.033}_{-0.006}$).
From this model, we obtain a black hole mass of $\log_{10}(M_{\rm BH}/M_\odot) = 7.58^{+0.33}_{-0.21}$.

The \civ\ emission line lag is $\tau_{\rm median} = 3.3^{+1.1}_{-0.7}$ days.
This is consistent with the \citet{agnstorm8} cross-correlation measurement of $\tau_{\rm cent} = 4.4 \pm 0.3$ days, measured using the same \civ\ emission line models.
We should note that they use a slightly longer campaign window ending at ${\rm THJD} = 6765$ rather than 6743, but this is unlikely to introduce a large change in the lag measurement.

Compared to the results of \citet{agnstorm9}, we find a smaller \civ\ BLR inclination angle ($\theta_i = 28.3^{+8.1}_{-9.2}$ degrees vs. $i=45$ degrees), but we note that \citet{agnstorm9} do not estimate uncertainties in their inclination angle fits.
We also find a stronger \civ\ response at shorter delays ($<5$ days) in our models.
This is evident in the velocity-integrated transfer function (Figure \ref{fig:transfer}, c, right panel) with the sharp peak in response at 1-2 days.

\subsection{\La\ (vs. UV light curve)} \label{sect:Lalpha}
\begin{figure}[h!]
\begin{center}
\includegraphics[width=3.45in]{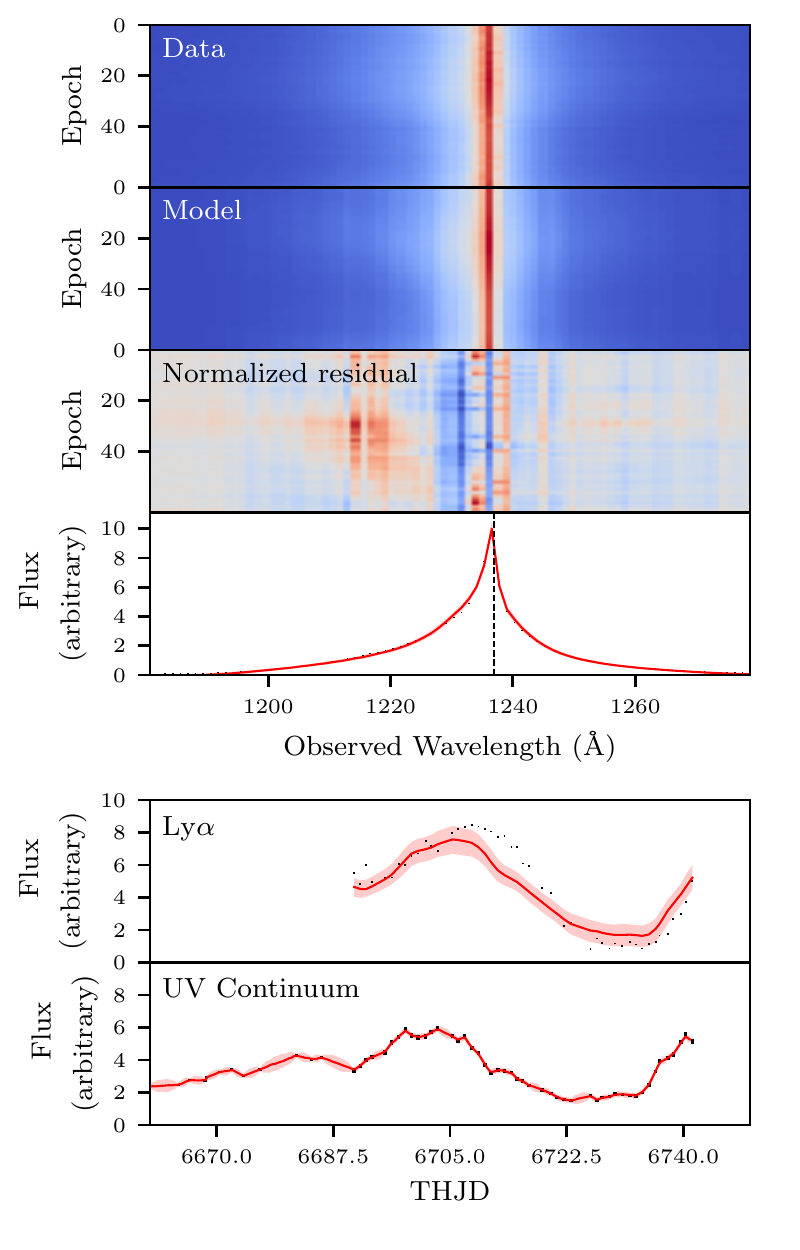}
\caption{Same as Figure \ref{fig:display_hb_uv}, but for the \La\ BLR models. The residuals at 1216~\AA\ are likely due to geocoronal \La\ emission.
\label{fig:display_lya}}
\end{center}
\end{figure}

As with \civ, we use the models from \citet{agnstorm8} for our \La\ data, binned by a factor of 32.
The model is able to fit the overall shape of the emission line quite well.
The overall shape of the emission line light curve is captured, but many of the models are unable to reproduce the amplitude of the emission line fluctuations (Figure \ref{fig:display_lya}, panel 5).
In order to fit the data without falling into local maxima in the likelihood space, we soften the likelihood with a temperature of $T=5000$, which is equivalent to increasing the uncertainties on the spectra by a factor of $\sqrt{5000} = 70.7$. 
Including such a high temperature allows us to measure realistic uncertainties on the model parameters.

\begin{figure}[h!]
\begin{center}
\includegraphics[width=3.3in]{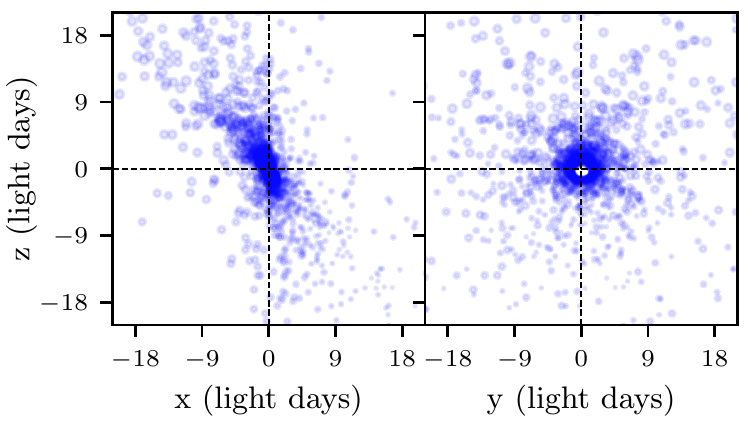}
\caption{
Same as Figure \ref{fig:geo_hb_uv}, but for the \La-emitting BLR.
\label{fig:geo_lya}}
\end{center}
\end{figure}

We find a \La\ BLR structure that is an inclined thick disk, with $\theta_o = 31.9^{+20.5}_{-12.2}$ degrees and $\theta_i = 23.7^{+23.6}_{-9.0}$ degrees (Figure \ref{fig:geo_lya}).
The radial distribution of particles drops off very quickly with radius, with $\beta = 1.86^{+0.10}_{-0.14}$.
The median radius of the BLR particles is $r_{\rm median} = 4.0^{+2.4}_{-1.7}$ light days, the minimum radius is $r_{\rm min} = 1.08^{+0.80}_{-0.49}$ light days, and the radial width is $\sigma_r = 23.3^{+15.3}_{-9.6}$ light days.
There is a small preference for emission back towards the ionizing source, with $\kappa = -0.23^{+0.52}_{-0.24}$.
There is little preference for the particles to be either uniformly distributed within the thick disk or located near the opening angles ($\gamma = 3.5^{+1.1}_{-1.5}$), nor is there a significant preference for either a transparent or opaque midplane ($\xi = -0.33^{+0.45}_{-0.25}$).

Dynamically, most of the particles are on either inflowing or outflowing trajectories ($f_{\rm ellip} = 0.20^{+0.16}_{-0.13}$), but it is not determined which direction of flow dominates ($f_{\rm flow} = 0.60^{+0.29}_{-0.40}$).
As with the models of the BLRs of the other lines, there is little contribution from macroturbulent velocities, with $\sigma_{\rm turb} = 0.018^{+0.049}_{-0.016}$.
The black hole mass based on the \La\ BLR models is $\log_{10}(M_{\rm bh}/M_{\odot}) = 7.38^{+0.54}_{-0.41}$

The models produce an emission line lag of $\tau_{\rm median} = 3.6^{+1.9}_{-1.7}$ days, which is consistent with the \citet{agnstorm8} cross-correlation measurement of $\tau_{\rm cent} = 4.8 \pm 0.3$ days.
Similar to \civ, we find a smaller \La\ BLR inclination angle than \citet{agnstorm9} ($\theta_i = 23.7^{+23.6}_{-9.0}$ degrees vs. $i=45$ degrees), but the values are still consistent due to the large uncertainty on our measurement and the lack of error bars by \citet{agnstorm9}.
We also find a shorter response than \citet{agnstorm9} for \La, with our model response peaking within 5 days, but the significance is difficult to asses without uncertainty estimates.

\section{Joint inferences on the BLR model parameters}
\label{sect:jointinference}

Ideally, our BLR model would reproduce all three emission lines and we would calculate the likelihood over all three data sets and adjust the model parameters for each region simultaneously.
Since we do not know which model parameters should be tied together, modeling each region individually provides a check on the consistency of the modeling method.
While the driving continuum used for each BLR model is the same, the spectra are all independent, and we can use the results from the three emission lines to put joint constraints on the model parameters.

\subsection{Black hole mass}
\label{sect:joint_mbh}

\begin{figure}[ht!]
\begin{center}
\includegraphics[width=3in]{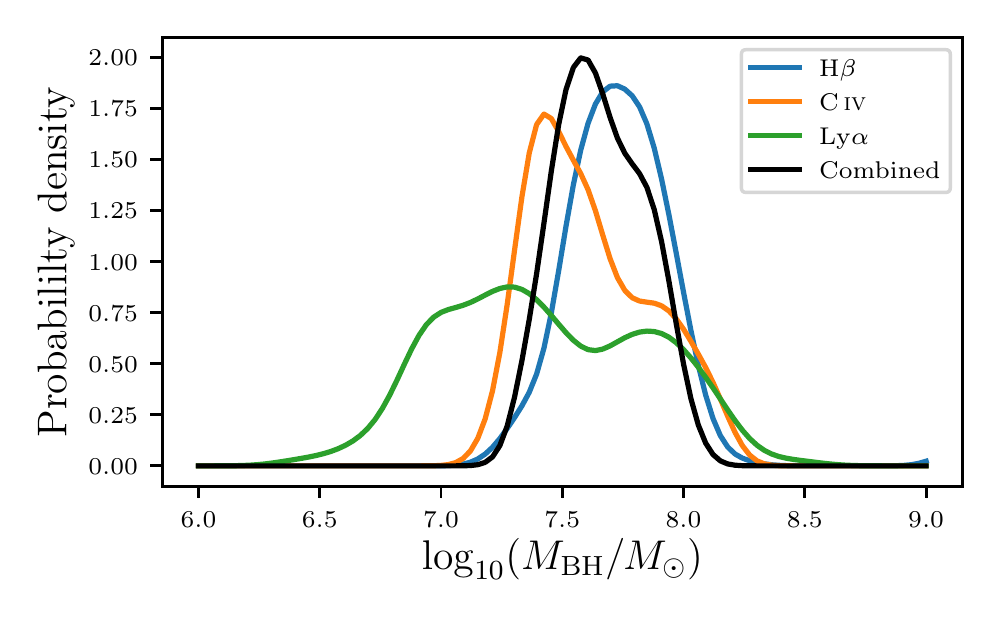} 
\caption{Joint inference on $\log_{10}(M_{\rm BH}/M_\odot)$ from combining the posterior PDFs for the three emission line region models.
\label{fig:joint_mbh}}
\end{center}
\end{figure}

Of all the BLR model parameters, we know that the black hole mass should be the same for all three emission lines.
Assuming that the three emission-line time series are independent, we can write
\begin{multline}
P(M_{\rm BH} | \dhb, \dciv, \dlya) = P(M_{\rm BH} | \dhb) \\
  P(M_{\rm BH} | \dciv) P(M_{\rm BH} | \dlya) / P(M_{\rm BH})^2,
\end{multline}
where $\dhb$, $\dciv$, $\dlya$ are the data for \Hb, \civ, and \La, respectively.
We use the \Hb\ BLR models fit with the UV continuum light curve so that the continuum data are the same for each emission line.
The BLR model uses a uniform prior in the log of $M_{\rm BH}$, so 
\begin{multline}
P[\log_{10}(M_{\rm BH}/M_\odot) | \dhb, \dciv, \dlya] \propto \\
  \prod_{i\in\{\hbm,\civm,\lam\}}P[\log_{10}(M_{\rm BH}/M_\odot) | \mathcal{D}_i].
\end{multline}
In practice, we estimate the posterior PDFs for the three emission lines using a Gaussian kernel density estimate (KDE) and multiply the three KDEs to obtain a joint constraint on the black hole mass.
The resulting joint posterior PDF is shown in Figure \ref{fig:joint_mbh}.
The individual $M_{\rm BH}$ measurements are all consistent with each other, and together provide a joint measurement of $\log_{10}(M_{\rm BH}/M_{\odot}) = 7.64^{+0.21}_{-0.18}$.

\begin{figure*}[h!]
\begin{center}
\includegraphics[width=6.8in]{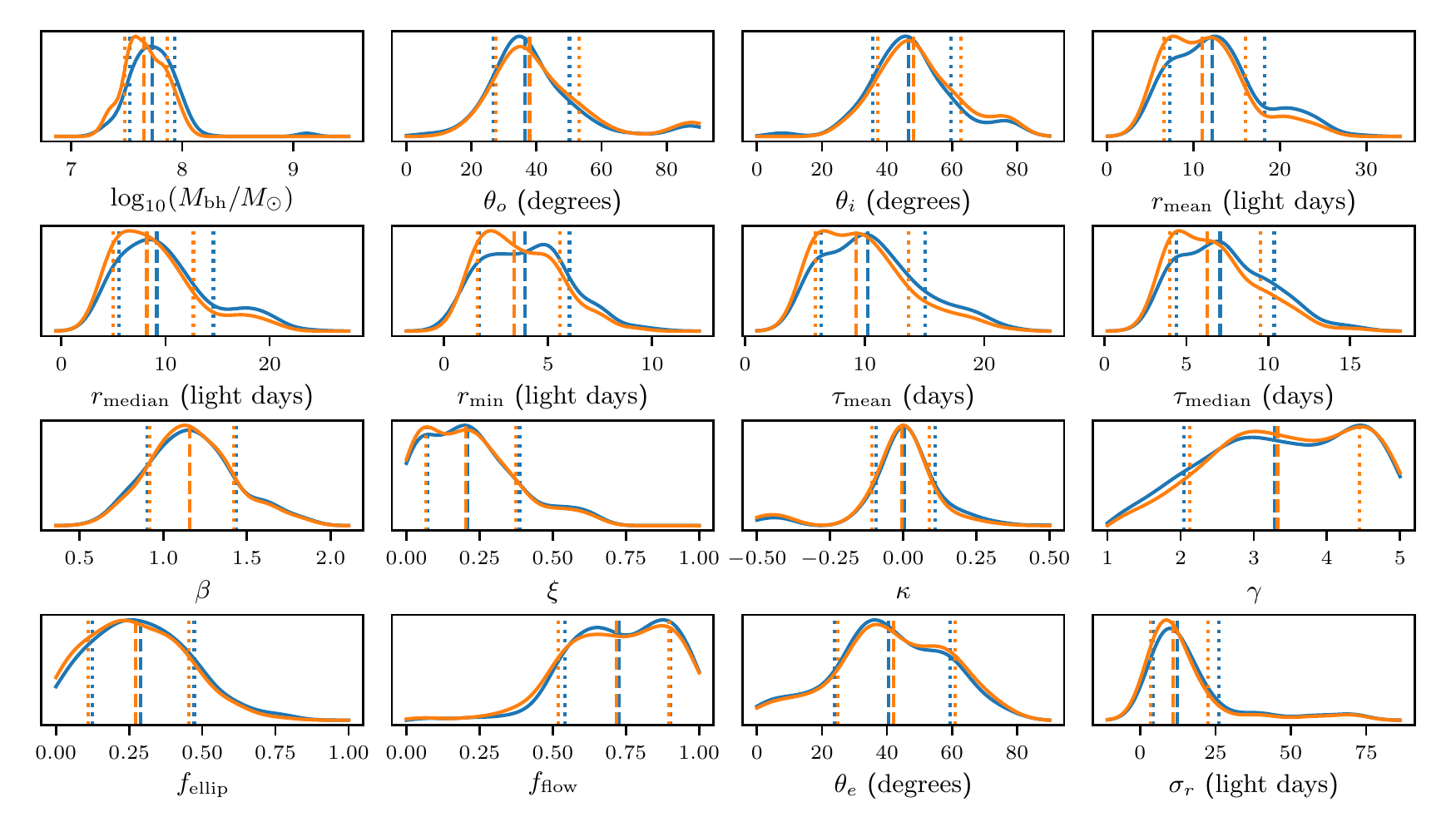} 
\caption{Gaussian KDE fits to the weighted (orange) and unweighted (blue) posterior PDFs for the \Hb\ BLR model parameters with the UV light curve as the driving continuum.
The weighting scheme used is the one described in Section \ref{sect:joint_mbh} in which the black hole masses for all three BLR models are forced to be the same.
The vertical dashed lines show the median value and the dotted lines show the 68\% confidence interval. 
\label{fig:posterior_weighted_hb}}
\end{center}
\begin{center}
\includegraphics[width=6.8in]{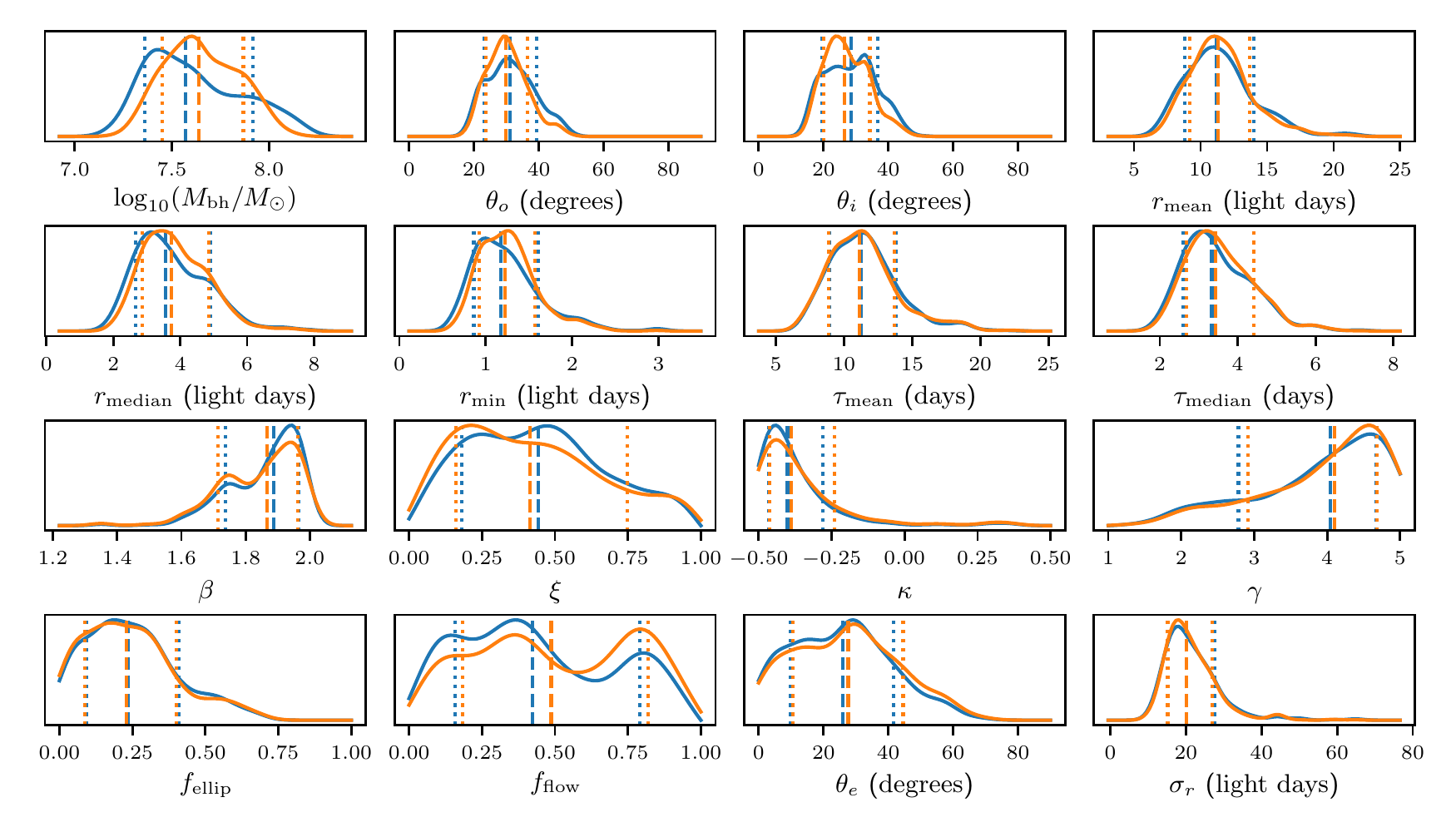} 
\caption{Same as Figure \ref{fig:posterior_weighted_hb}, but for the \civ\ BLR models. The weighting scheme used is the one described in Section \ref{sect:joint_mbh} in which the black hole masses for all three BLR models are forced to be the same.
\label{fig:posterior_weighted_civ}}
\end{center}
\end{figure*}

\begin{figure*}[ht!]
\begin{center}
\includegraphics[width=6.8in]{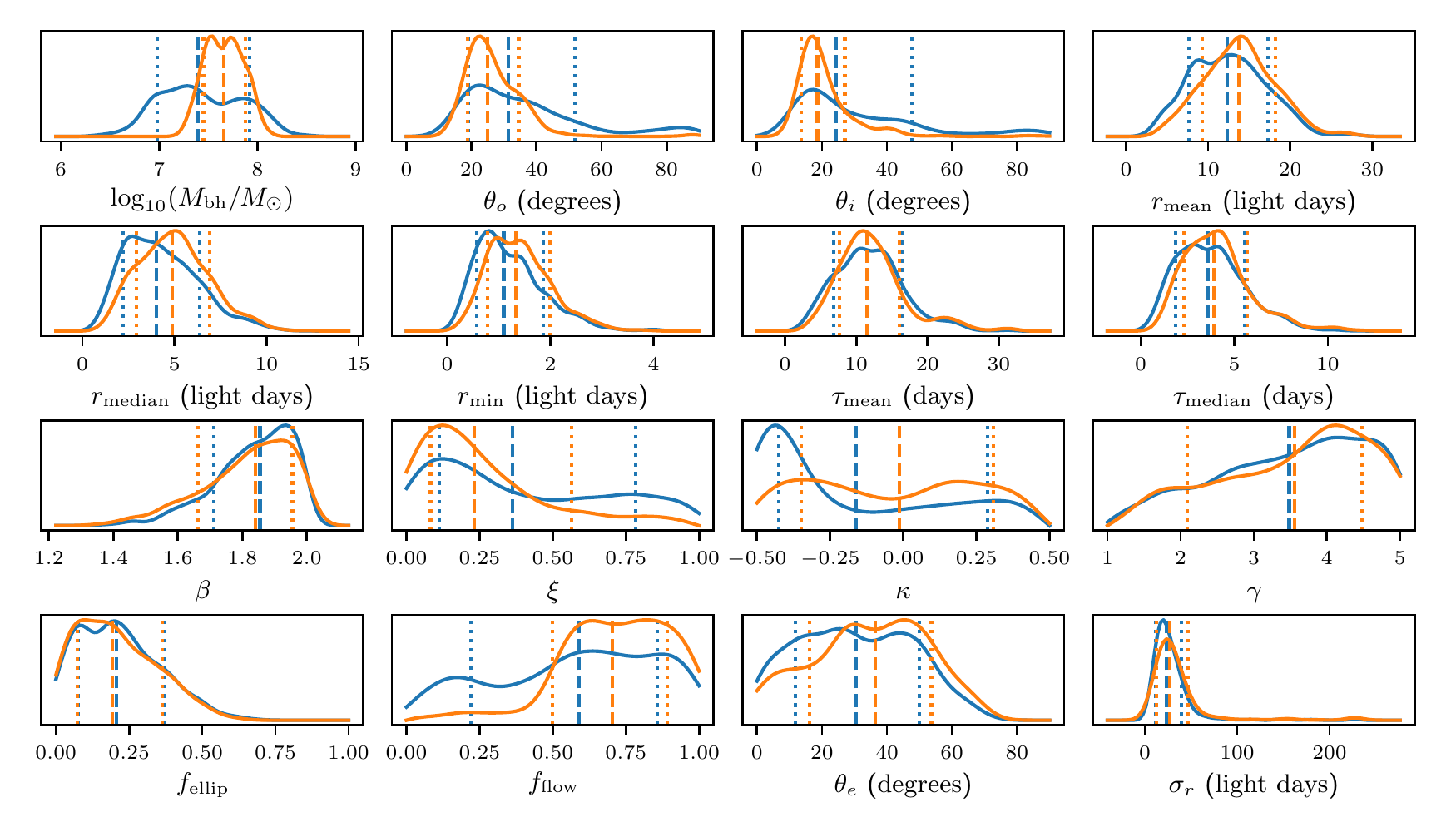} 
\caption{Same as Figure \ref{fig:posterior_weighted_hb}, but for the \La\ BLR models. The weighting scheme used is the one described in Section \ref{sect:joint_mbh} in which the black hole masses for all three BLR models are forced to be the same.
\label{fig:posterior_weighted_lya}}
\end{center}
\end{figure*}

Using our joint constraint on the black hole mass, we can use the method of importance sampling \citep[see, e.g., ][]{Lewis+02} to further constrain the other parameters of our BLR models.
Importance sampling is a technique that allows one to sample an unavailable distribution $P_2$ via a distribution $P_1$ that can be more easily sampled.
By writing $P_2 = (P_2/P_1) P_1$, we simply need to determine the weighting factor $P_2/P_1$.
In our case, $P_2$ is the posterior PDF for the BLR parameters for, say, \Hb, given all emission line data:
\begin{align}
P_2 = P(\theta_{\hbm}, M_{\rm BH} | \dhb, \dciv, \dlya);
\end{align}
and $P_1$ is the posterior PDF given only the \Hb\ data:
\begin{align}
P_1 = P(\theta_{\hbm}, M_{\rm BH} | \dhb).
\end{align}
Here, $\theta_{\hbm}$ are the \Hb\ BLR model parameters \textit{not} including the black hole mass.
The weight $P_2/P_1$ is simply the ratio of our joint PDF on $M_{\rm BH}$ to the PDF based on the individual lines.

The result of this method is that the posterior samples with $M_{\rm BH}$ in regions of high density in the joint PDF will be weighted higher than those with $M_{\rm BH}$ in regions of lower density.
This can be useful to exclude regions of parameter space that might fit the emission-line time series well, but with an incorrect black hole mass.
Gaussian KDE fits to the original and importance sampled posterior PDFs are shown in Figures \ref{fig:posterior_weighted_hb} - \ref{fig:posterior_weighted_lya}.

Examining the weighted results, we find little change to the \Hb\ BLR parameters, other than a slight decrease in the parameters indicating the size of the BLR.
The joint constraint on the black hole mass is slightly lower than the individual \Hb\ constraint, so this results in preferring BLR geometries that are slightly smaller.
The \civ\ BLR parameters also show almost no change.
The posterior PDFs for the \La\ BLR parameters show the largest change due to the largest difference between the \La-only $M_{\rm BH}$ PDF  and the joint PDF.
The solutions with low $M_{\rm BH}$ are essentially excluded, resulting in a very slight increase in radius, and a more robustly determined low inclination angle.
Additionally, the kinematics go from being relatively undetermined towards a preference for outflow.

\subsection{Black hole mass and inclination angle}
\label{sect:joint_mbh_thetai}

\begin{figure}[h!]
\begin{center}
\includegraphics[width=3.4in]{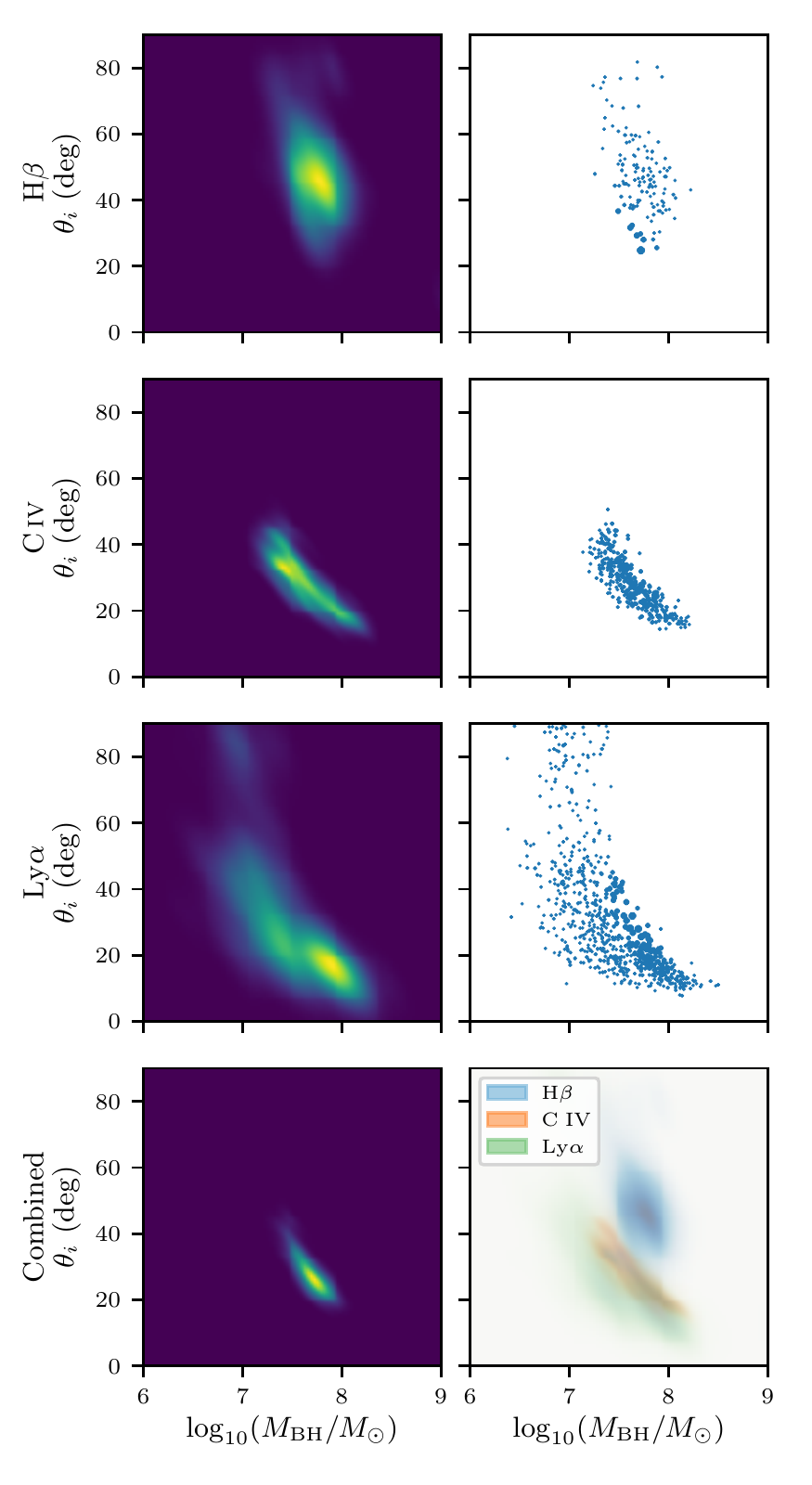}
\caption{\textit{Left}: Gaussian KDEs for the 2d posterior PDFs for $(\log_{10}(M_{\rm BH}/M_\odot), \theta_i)$ in each BLR model as well as the joint constraint (bottom). \textit{Right}: Weighted posterior samples for the three BLR models (top 3), and the region of overlap of the PDFs in the left column (bottom). The size of each point corresponds to the sample's weight. The weighting scheme used is the one described in Section \ref{sect:joint_mbh_thetai} in which the black hole masses and inclination angles for all three BLR models are forced to be the same.
\label{fig:joint_mbh_thetai}}
\end{center}
\end{figure}

We can also examine the scenario in which both the black hole mass and the inclination angle are assumed to be the same for each line-emitting region.
We follow the same methods discussed in Section \ref{sect:joint_mbh}, except in this case we examine the 2D posterior PDF for $(\log_{10}(M_{\rm BH}/M_\odot), \theta_i)$.
Figure \ref{fig:joint_mbh_thetai} shows the Gaussian KDE fits to the 2D posterior PDFs, as well as the joint posterior PDF.
From the figure, we see that there is little overlap between the \Hb\ model parameters and the \civ\ and \La\ model parameters.
Thus, when we calculate the weights to importance sample the \Hb\ BLR posterior PDFs, only a very small portion of the parameter space receives a significant weight.

\begin{figure*}[ht!]
\begin{center}
\includegraphics[width=6.8in]{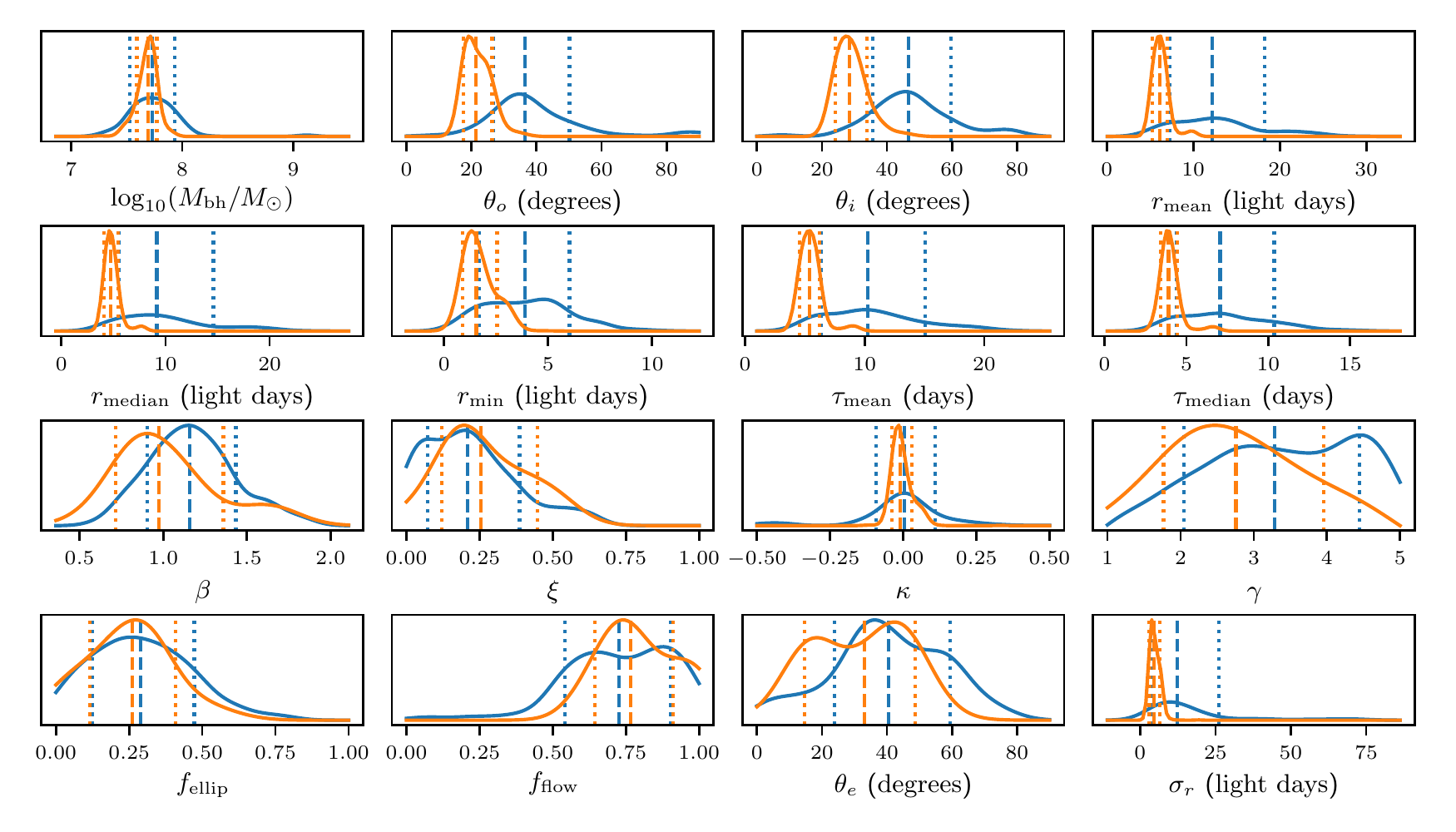}
\caption{Same as Figure \ref{fig:posterior_weighted_hb}, for the \Hb\ BLR models, but when both $M_{\rm BH}$ and $\theta_i$ are enforced to be the same as those inferred by the \La\ and \civ\ models (described in Section \ref{sect:joint_mbh_thetai}).
\label{fig:posterior_weighted_thetai_hb}}
\end{center}
\end{figure*}

Examining the weighted posterior PDFs in Figure \ref{fig:posterior_weighted_thetai_hb}, we see that only models with extremely small \Hb\ BLRs are not excluded.
In fact, for the \Hb\ BLR inclination angle to match that of \civ \ and \La, the \Hb-emitting BLR would need to be \textit{smaller} than the \civ- and \La-emitting BLRs.
This directly contradicts the plentiful studies showing ionization stratification within the BLR \citep[e.g.,][]{clavel91, reichert94}.
Additionally, this would require an \Hb\ lag of $\tau_{\rm median} = 3.9^{+0.5}_{-0.5}$ days, which is significantly shorter than the measurements of $\tau_{\rm cen,T1} = 7.62^{+0.49}_{-0.49}$ days and $\tau_{\rm JAVELIN, T1} = 6.91^{+0.64}_{-0.63}$ days by \citet{agnstorm5}.
Given these contradictions as well as the clear offset in the $(\log_{10}(M_{\rm BH}/M_\odot), \theta_i)$ posterior PDFs, we conclude that the assumption of identical $\theta_i$ must be faulty.

\section{Discussion}
\label{sect:discussion}

\subsection{Effect of the continuum light curve choice on modeling results}
For most reverberation mapping data sets suitable for dynamical modeling, the only continuum light curve we have access to is the optical light curve, so we treat this as a proxy for the ionizing continuum light curve.
In reality, these are not the same light curves and arise in different locations both in space and time. 
The optical continuum light curve is a delayed and smoothed version of the ionizing continuum light curve with an additional contribution from diffuse continuum emission, so short time scale variability information is lost.
The UV continuum is closer to the ionizing continuum, and is thus closer to the assumptions of our model.
With these data, we have access to both light curves, so we can examine how the choice of continuum affects the modeling results.

Figure \ref{fig:posterior_hb_comparison} shows the model parameter posterior PDFs for the two versions plotted on top of each other.
Comparing the two sets of results, we find that the continuum light curve choice primarily affects the parameters dictating the scale of the BLR, but not the parameters that describe the shape.
The median radius of the BLR is found to be roughly 3 light days smaller when the $V$-band light curve is used instead of the UV light curve, although the results still agree to within the uncertainties.
Similarly, the minimum radius is 1.5 light days smaller, but is again in agreement to within the uncertainties.
\citet{agnstorm3} measure a 1.86 day lag between the \emph{HST} $\lambda$1157.5~\AA\ and $V$-band light curves, which is consistent with the differences in the BLR model size parameters.

Since the black hole mass measurement depends on the scale of the BLR, it is important to note that this parameter will be affected by the choice of the continuum light curve.
In black hole mass measurements based on the use of the scale factor $f$, this issue is mitigated by the fact that $f$ itself is calibrated using the same light curves that exhibit the delay \citep[e.g.,][]{onken04, collin06, woo10, woo13, grier13b, batiste17}.
Since the dynamical modeling approach treats the black hole mass directly as a free parameter, the under-estimate of the BLR size leads to under-estimating the black hole mass.
In particular, $M_{\rm BH}$ as measured by the model with the $V$-band light curve should be smaller than that measured with the UV light curve by a factor of $\tau_{V}/\tau_{\rm UV}$, where $\tau_{V}$ ($\tau_{\rm UV}$) is the lag between the $V$-band (UV) continuum fluctuations and emission line fluctuations.
For this data set, this is a factor of $\sim$$2/3$ ($0.18$ in $\log_{10}[M_{\rm BH}/M_{\odot}]$), which is consistent with our model masses.
However, NGC 5548 deviated significantly from the typical $r_{\rm BLR} - L_{\rm AGN}$ relation during this campaign, with an \Hb\ BLR size smaller than expected by a factor of $\sim5$ \citep{agnstorm5}.
It is possible that for most AGN, the BLR is significantly larger than $c \times \tau_{V}$ so that $\tau_{\rm UV}/\tau_{V}$ is closer to unity and the effect of using the $V$ band as a proxy is mitigated.
Unfortunately, the UV-optical lag is typically not available for the campaigns in which the $V$ band is used, which makes finding a $M_{\rm BH}$ correction factor complicated.
Further research will be required to understand how to make such corrections to models of these data.

We should also note that based on the \Hb\ BLR size and the UV-optical lag, the optical light curve we measure arises in a region that is spatially extended as seen by the BLR.
However, this alone does not significantly affect the point-like continuum assumption of our model as long as the true ionizing source is still close to point-like. 
Rather, the only effects are the shortened time-lags discussed above and a smoothing of features in the continuum light curve.
Reassuringly, we find that no other parameters in the BLR model are affected.

\subsection{Comparison with previous \Hb\ modeling}

NGC 5548 was also monitored as part of the Lick AGN Monitoring Project 2008 \citep[LAMP,][]{walsh09}, and those data were modeled using the same code as in this paper.
The AGN was at a lower luminosity state during the LAMP 2008 campaign, with a host-galaxy + AGN flux density of $f_{\lambda}[5100 \times (1 + z)] = 6.12 \pm 0.38 \times 10^{-15}\rm{~erg~s}^{-1}\rm{~cm}^{-2}\rm{~\AA}^{-1}$ \citep{bentz09}.
Comparatively, \citet{agnstorm5} measure $F_{5100,{\rm total}} = 11.31 \pm 0.08 \times 10^{-15}\rm{~erg~s}^{-1}\rm{~cm}^{-2}\rm{~\AA}^{-1}$ for the portion of the campaign before the BLR holiday.
While the exact host-galaxy correction depends on the slit sizes and position angles for the two campaigns, the $f_{\lambda,{\rm gal}}[5100 \times (1 + z)] = 3.752 \pm 0.375 \times 10^{-15}\rm{~erg~s}^{-1}\rm{~cm}^{-2}\rm{~\AA}^{-1}$ measurement from \citet{bentz13} means that the AGN was roughly 4 times brighter in 2014 than in 2008.
From the $r_{\rm BLR}-L$ relation \citep[e.g.,][]{bentz13}, we would expect the BLR size to be smaller during the LAMP 2008 campaign than in the 2014 campaign by a factor of $\sim$2.

\citet{pancoast14b} found a BLR structure in NGC 5548 that was also an inclined thick disk with $\theta_o = 27.4^{+10.6}_{-8.4}$ degrees and $\theta_i = 38.8^{+12.1}_{-11.4}$ degrees.
The mean and minimum radii were $r_{\rm mean} = 3.31^{+0.66}_{-0.61}$ and $r_{\rm min} = 1.39^{+0.80}_{-1.01}$ light days, respectively, and the radial width was $\sigma_r = 1.50^{+0.73}_{-0.60}$ light days.
They found a radial distribution between exponential and Gaussian with $\beta = 0.80^{+0.60}_{-0.31}$ and a spatial distribution described by $\gamma = 2.01^{+1.78}_{-0.71}$.
Finally, they found a preference for emission back towards the ionizing source ($\kappa = -0.24^{+0.06}_{-0.13}$) and a mid-plane that is mostly opaque ($\xi = 0.34^{+0.11}_{-0.18}$).

Dynamically, they find a BLR that is mostly inflowing ($f_{\rm flow} = 0.25^{+0.21}_{-0.16}$) with the fraction of particles on elliptical orbits only $f_{\rm ellip} = 0.23^{+0.15}_{-0.15}$.
Of the inflowing orbits, most are bound with $\theta_e = 21.3^{+21.4}_{-14.7}$ degrees.
They do not find a significant contribution from macroturbulent velocities ($\sigma_{\rm turb} = 0.016^{+0.044}_{-0.013}$).
The black hole mass \citet{pancoast14b} measure is $\log_{10}(M_{\rm BH}/M_\odot) = 7.51^{+0.23}_{-0.14}$.

\begin{figure*}[h!]
\begin{center}
\includegraphics[width=6.8in]{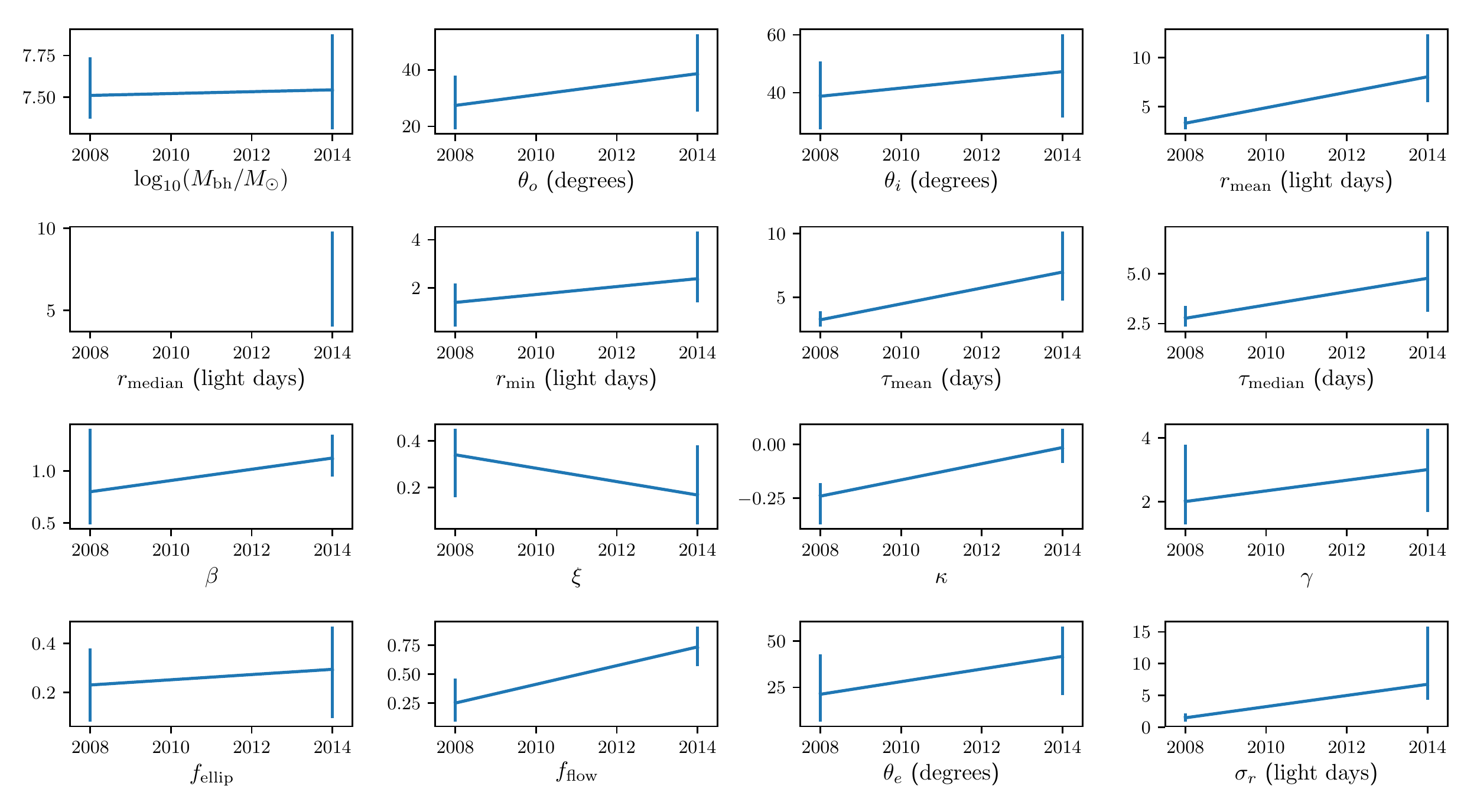}  
\caption{Change in the \Hb\ BLR model parameters from the LAMP 2008 campaign to the AGN STORM campaign. The vertical bars indicate the posterior PDF median and 68\% confidence intervals for the two campaigns. The value of $r_{\rm median}$ is missing for the 2008 campaign since it was not reported by \citet{pancoast14b}.
\label{fig:lamp_storm_comparison}}
\end{center}
\end{figure*}

Figure \ref{fig:lamp_storm_comparison} shows the change in model parameters from \citet{pancoast14b} and the \Hb\ vs. $V$ band modeling results from this paper.
As expected, the parameters describing the size of the BLR increase from the 2008 campaign to the 2014 campaign.

Other parameters that changed from the 2008 campaign and 2014 campaign were $f_{\rm flow}$ and $\kappa$.
The change in $f_{\rm flow}$ indicates a switch from net-inflowing gas to net-outflowing gas.
If true, this could suggest a significant change in the kinematics of the broad-line region that might be connected with the increase in AGN luminosity.
However, we should note that with $\theta_e = 42^{+16}_{-21}$ degrees for the AGN STORM campaign, the outflowing particles could be on highly elliptical bound orbits rather than on pure radial outflowing trajectories.
The parameter $\kappa$ showed a preference for \Hb\ emission from BLR clouds back towards the ionizing source in the 2008 campaign, but indicates a preference for isotropic emission in this data set.

Reassuringly, the black hole mass, opening angle, and inclination angles all remain consistent for the two data sets, as we would not expect these to change on a six-year timescale.
Additionally, $\xi$ remains the same, indicating a mostly opaque mid-plane.
The parameters $\gamma$ and $\sigma_{\rm turb}$ were not well constrained in either the 2008 or 2014 campaign models.
Finally, the $\beta$ parameter of the Gamma distribution was poorly constrained with the 2008 campaign data but is better determined with the 2014 campaign data.

This comparison of modeling results of a single AGN over multiple campaigns represents the second of its nature, with the first being Arp 151, presented by \citet{Pancoast++18}.

\subsection{Comparison of the three line-emitting regions}
The AGN STORM data set is the first data set in which this modeling technique can be applied to multiple emission lines for the same AGN.
This gives us a unique opportunity to examine how the structure and kinematics of the three line-emitting regions are the same and how they differ.
In Figure \ref{fig:linecomparison}, we compare the posterior PDFs for the three BLR models.
Each model used the same UV light curve as the driving continuum.

\begin{figure}[h!]
\begin{center}
\includegraphics[width=3.3in]{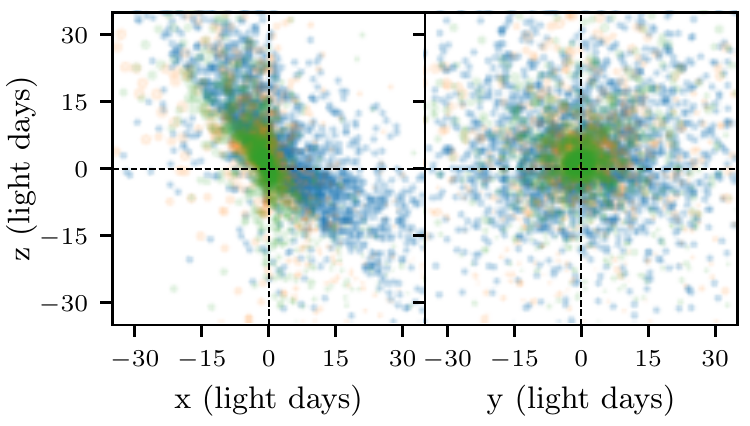}  
\includegraphics[width=3.3in]{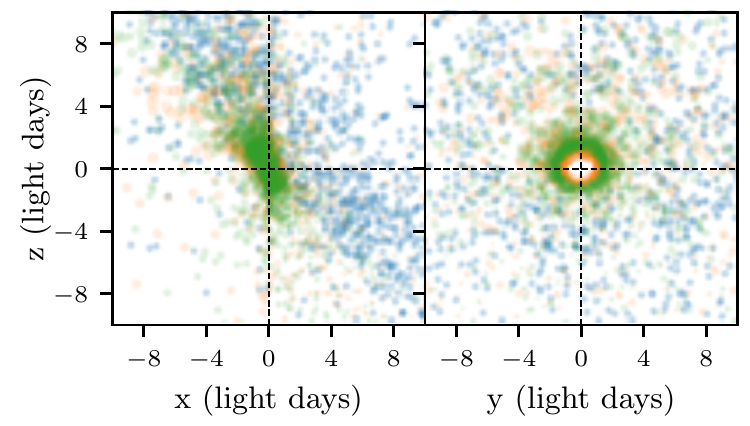}  
\includegraphics[width=3.3in]{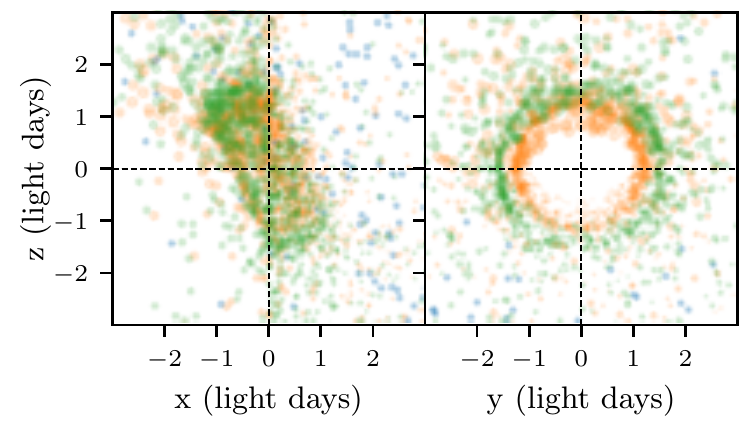}  
\caption{Possible geometries of the three line-emitting regions, with \Hb, \civ, and \La\ in blue, orange, and green, respectively. All panels show the same three geometries from different angles and different distance scales. Note that each model displayed is only one possible model from the posterior distribution, selected to have parameters closest to the median values reported in Table \ref{tab:results}, and the exact radial ordering of \civ\ and \La\ is not constrained.
\label{fig:geo_compare}}
\end{center}
\end{figure}

Examining the differences in model parameters, we clearly see radial ionization stratification (see, e.g., the $r_{\rm median}$ distributions).
Additionally, the radial distribution of particles is significantly different, with the \civ\ and \La\ BLRs having $\beta$ close to 2 while the \Hb\ BLR has $\beta\sim 1$.
This also becomes clear when we show possible geometries of the three BLRs plotted on top of each other in Figure \ref{fig:geo_compare}.
There is clear radial structure in the three line-emitting regions, with \civ\ and \La\ emission coming from a very localized portion of a shell, while the \Hb\ region is much more spread out in the radial direction.
The models displayed in the figure show the \civ\ BLR with a smaller minimum radius than the \La\ BLR, but the ordering of these two lines is not well constrained by the posterior parameter distributions.

While the $r_{\rm min}$ parameter is not well constrained for the \Hb~vs.~UV models, the median value suggests that there is a $\sim$2.5 light day region between $r_{{\rm min, Ly}\alpha}$ and $r_{{\rm min, H}\beta}$ in which there is \La\ emission but no \Hb\ emission.
It is likely that there is still \Hb\ emission in this region, but in order to fit the stronger emission at larger wavelengths, the $r_{\rm min}$ parameter is shifted to larger radii.
We discuss the possibility of tying the line emission to the underlying BLR gas distribution in Section \ref{sect:underlying_gas}.

The opening angle is surprisingly consistent between the three line-emitting regions.
The inclination angle, on the other hand, shows some discrepancy.
While it does not appear to be a huge difference, the discrepancy is at the $>1\sigma$ level, with $\theta_i = 46.1^{+13.4}_{-9.0}$ (\Hb), $\theta_i = 28.3^{+8.1}_{-9.2}$ (\civ), $\theta_i = 23.7^{+23.6}_{-9.0}$ (\La).
We examine this further in Section \ref{sect:joint_mbh_thetai} and find that enforcing $\theta_i$ to be equal for all three regions leads to unphysical results in the radial ionization stratification of the BLR.
Given that the \Hb\ BLR extends to a much larger radius than the \civ\ and \La\ BLRs, it is possible that they may lie at slightly different inclinations.
For instance, a warped disk geometry would show a different inclination angle near the center than at larger radii.
Since our model does not fit the underlying BLR gas, it is unclear if the discrepancy arises from the gas distribution itself or is an effect only present in the gas emission.

\subsection{Systematic Uncertainties and Model Limitations}
\subsubsection{A simple physical model}

When interpreting the results, it is important to keep in mind that we are using a simple model to describe what is likely a very complex region of gas.
The current implementation of the code is not intended to explain the exact physical processes within the BLR, but rather to describe the overall size and shape of the BLR emission.
It would be computationally infeasible to explore the parameter space of a full physical model of the BLR, so we neglect the details of, e.g., photoionization physics and radiation pressure and instead use a simple, flexible model that is designed to account for a wide range of possible BLR gometries and kinematics, while keeping the number of parameters and computational speed tractable.
While these simplifications allow us to constrain the overall BLR structure and velocity field, there are certain details of the BLR that go un-modeled \citep[see][Section 2.2 for a discussion]{Raimundo++19b}.

A blind test of reverberation mapping techniques found that for a mock data set, the inferred model parameters were in excellent agreement with the input BLR model, even though the details of the transfer function and RMS profile were not fully captured \citep{Mangham++19}.
Efforts are currently underway (Williams et al., in prep) to include a more physically realistic description of the photoionization physics in the BLR.
These additions to the model will provide the flexibility to fit more variability features in the emission line, and will naturally allow for effects such as ``breathing'' of the BLR.

\subsubsection{Correlations among model parameters}
With the high dimensionality of the BLR model parameterization comes a number of correlations between the model parameters.
\citet{Grier++17} discuss in detail a degeneracy between the opening angle and inclination angle, pushing these two parameters towards similar values.
In essence, in order to produce the single-peaked emission-line profiles we observe, $\theta_o \gtrsim \theta_i$, effectively putting a prior on the opening angle from $\theta_i$ to $90$ degrees.
Therefore, it is possible that the BLRs actually have $\theta_o < \theta_i$, but have a structure and kinematics that cannot be reproduced by the current version of the model.

Additionally, given the parameterization of the model, there are multiple ways to combine model parameters to produce the same BLR model.
For instance, as $\theta_e\rightarrow 90$ degrees, nearly all particles are placed in near-circular orbits, regardless of the value of $f_{\rm ellip}$ or $f_{\rm flow}$.
Similarly, a model with $\theta_i,\theta_o \rightarrow 90$ degrees and $\gamma\rightarrow 5$ produces a line of particles perpendicular to the observer's line of sight.
However, this is equivalent to a face-on disk since rotations in the plane of the sky cannot be resolved with reverberation mapping data.
These situations can increase the uncertainty on individual model parameters even if the particle distributions are very well determined.

\subsubsection{Emission line model}
When modeling a BLR, we assume that we can accurately isolate the broad emission line from contaminant features in the region of the line.
If the contaminants are left in, the model will try to compensate by adjusting the parameters to fit this extra emission.
\citet{Williams++18} show that the choices made when modeling an emission line, such as choice of \feii\ template, may influence the line profile enough to have an effect on the resulting model parameters.
\citet{agnstorm5} discuss the issues in decomposing the optical spectra for NGC 5548, including degeneracies between weak \feii\ and the continuum light as well as weak \hei\ emission blended with \Hb.
Similarly, the \La\ and \civ\ raw spectra have significant amounts of broad and narrow absorption which must first be modeled, making our resulting BLR models inherently dependent on the emission line models.

\subsubsection{Underlying BLR gas}
\label{sect:underlying_gas}
It is important to understand that the model use in this work is fitting the BLR gas {\it emission} and not the gas itself.
There is, of course, gas elsewhere in the BLR that we do not see either because it is not emitting or because the emission is obscured.
For instance, the fact that we see \La\ emission within $r_{{\rm min, H}\beta}$ shows that emitting hydrogen gas is present in this region, yet we are unable to detect sufficiently strong \Hb\ emission.

Given a distribution of gas around the central BH and an ionizing spectrum, photoionization calculations are able to predict line emissivities through the BLR.
Future dynamical modeling implementations can use these calculations to determine the distribution and motions of the {\it underlying gas} in the BLR, as well as the line emission.
This will help shed light on some of the effects we see, such as the different inclination angles for \civ, \La, and \Hb\ emission.

Although the model used here does not have these features, its current aim is not to provide a full physical description of the BLR.
Rather, we wish to describe the overall structure and motions of the BLR emission, and use this as a tool to measure black hole masses.
Despite its limitations, the simple model achieves these goals, as evidenced by the consistent black hole mass measurements, agreement with cross-correlation lag measurements, and similar geometries to those inferred from the velocity-delay maps of \citet{agnstorm9}.


\section{Summary}
\label{sect:summary}

We have fit dynamical models of the BLR to three emission lines using the AGN STORM data set.
This is the first time the modeling approach has been used to fit multiple emission lines for the same AGN, and is the first time it has been used with UV emission lines.
Additionally, we fit the \Hb\ emission-line time series using both the UV light curve and $V$-band light curve as the driving continuum.
This has allowed us to better understand the systematics involved in other modeling results when only the optical continuum is available (e.g., ground-based campaigns).

The main results of our analysis can be summarized as follows:
\begin{enumerate}
\item Modeling of \Hb, \civ, and \La\ provides three independent black hole mass measurements that are in good agreement.
A joint inference combining all three lines gives $\log_{10}(M_{\rm BH}/M_{\odot}) = 7.64^{+0.21}_{-0.18}$.
This is consistent with cross-correlation- and {\sc MEMEcho}-based measurements with these data.
\item Based on the model, we infer a radial structure in the BLR, with \civ\ and \La\ emission arising at smaller radii than \Hb.
The corresponding lags for our models are consistent with the cross-correlation and {\sc JAVELIN} measurements of \citet{agnstorm5} and \citet{agnstorm8}.
\item The different line-emitting regions do not need to lie in the same inclination plane.
In NGC 5548, the \civ\ and \La\ BLRs share the same inclination angle, while the more extended \Hb\ BLR lies at a slightly higher inclination.
\item When the optical light curve is used as the driving continuum, the model parameters describing the \Hb\ BLR size ($r_{\rm mean}$, $r_{\rm median}$, $r_{\rm min}$) are smaller by an amount comparable to the UV-optical lag, as opposed to when the UV light curve is used, and the black hole mass is under-estimated by a factor of $\tau_V/\tau_{\rm UV}$.
The parameters describing the BLR geometry and kinematics, however, are not significantly affected.
This indicates that the $V$-band continuum is a suitable proxy for the ionizing continuum when studying the BLR structure and kinematics, but the UV-optical lag must be considered when measuring the BLR size.
\item The radius of the \Hb-emitting BLR increased by a factor of $\sim$3 between the 2008 LAMP campaign and the 2014 AGN STORM campaign, but the measured black hole mass remained constant.
The other geometric parameters remained consistent in this time frame. 
There may have been a change in the BLR kinematics from inflow to outflow, although this is not robustly determined.
\end{enumerate}

With the exquisite data analyzed in this paper, we have challenged the modeling method to recover the same black hole mass given three sets of data and to provide BLR properties using multiple light curves as the driving continuum.
The consistent results have demonstrated that the modeling approach is a robust method of determining the BLR structural and kinematic properties, and reliable black hole mass measurements can be extracted from \La\ and \civ\ in addition to \Hb.
Further, we have shown that the $V$-band continuum is a suitable proxy for the ionizing continuum for measuring BLR structural and kinematic properties, and reliable black hole mass estimates can be made provided the UV-optical lag is accounted for.
The findings have provided insights into how the different line-emitting portions of the BLR fit together and how they evolve over time, and will help inform future improvements to the BLR model.


\acknowledgments  
Support for \textit{HST} program No. GO-13330 was provided by NASA through a grant from the Space Telescope Science Institute, which is operated by the Association of Universities for Research in Astronomy, Inc., under NASA contract NAS5-26555.
Research by P.R.W. and T.T. is supported by NSF grants AST-1412315 and AST-1907208.
T.T. and P.R.W. acknowledge support from the Packard Foundation through a Packard Fellowship to T.T.
Research by AJB was supported by NSF grant AST-1907290.
G.F. and M.D. acknowledge support by NSF (1816537, 1910687), NASA (17-ATP17-0141, 19-ATP19-0188),
and STScI (HST-AR-15018, HST-AR-14556).
M.I. acknowledges the support form the NRF grant, No. 2020R1A2C3011091, funded by the Korea government (MSIT).
M.V. gratefully acknowledges support from the Independent Research Fund Denmark via grant number DFF 8021-00130.
C.S.K. is supported by NSF grants AST-1908952 and AST-1814440.
V.N.B. acknowledges assistance from a NASA grant associated with HST proposal GO 15215, a NASA ADAP grant (grant \# 80NSSC19K1016) and a National Science Foundation (NSF) Research at Undergraduate Institutions (RUI) grant (AST-1909297).  Note that findings and conclusions do not necessarily represent views of the NSF.
D.J.S. acknowledges funding support from the Eberly Research Fellowship from The Pennsylvania State University Eberly College of Science.  The Center for Exoplanets and Habitable Worlds is supported by the Pennsylvania State University, the Eberly College of Science, and the Pennsylvania Space Grant Consortium.
A.V.F.'s group at U.C. Berkeley is grateful for the financial support of NSF grant AST-1211916, the TABASGO Foundation, the Christopher R. Redlich Fund, and the Miller Institute for Basic Research in Science (U.C. Berkeley). Research at Lick Observatory is partially supported by a generous gift from Google.

\bibliographystyle{apj}
\bibliography{references}

\end{document}